\documentclass[aps,twocolumn,prc,superscriptaddress,nofootinbib,floatfix]{revtex4-2}

\usepackage{graphicx}
\usepackage{mathtools,amssymb,bm}
\usepackage{upgreek}
\usepackage{booktabs}
\usepackage{braket}
\usepackage{multirow}
\usepackage{physics}
\usepackage[version=3]{mhchem}
\usepackage[table,dvipsnames]{xcolor}
\usepackage[colorlinks=true]{hyperref}

\newcommand{\thor}{$^{229}$Th}

\newcommand{\lisaf}{LiSrAlF$_6$}

\newcommand{\caf}{CaF$_2$}

\begin{document}

\title{{\em Ab initio} calculations of \ce{^{229}Th} band-to-band internal conversion rate in  \ce{^{229}ThO2}}

\author{Udeshika C. Perera}
\affiliation{Department of Physics, University of Nevada, Reno, Nevada 89557, USA}
\author{H. B. Tran Tan}
\affiliation{Computational Physics Division, Los Alamos National Laboratory, P.O. Box 1663, Los Alamos, New Mexico 87545, USA} 
\author{H. W. T. Morgan}
\affiliation{Department of Chemistry, University of Manchester, Oxford Road, Manchester M13 9PL, UK}
\author{Eric R. Hudson}
\affiliation{Department of Physics and Astronomy, University of California Los Angeles, Los Angeles, CA, USA}
\affiliation{Challenge Institute for Quantum Computation, University of California Los Angeles, Los Angeles, CA, USA}
\affiliation{Center for Quantum Science and Engineering, University of California Los Angeles, Los Angeles, CA, USA}
\author{Daniel A. Rehn}
\affiliation{Computational Physics Division, Los Alamos National Laboratory, P.O. Box 1663, Los Alamos, New Mexico 87545, USA} 
\author{Andrei Derevianko }
\affiliation{Department of Physics, University of Nevada, Reno, Nevada 89557, USA}
\date{\today}

\begin{abstract}
We present an \textit{ab initio} calculation of the band-to-band internal-conversion rate of the $\hbar \omega_\mathrm{nuc} \approx 8.35~\mathrm{eV}$ isomeric transition in \ce{^{229}ThO2}. Because the nuclear transition energy exceeds the electronic band gap of \ce{ThO2}, the isomer can decay nonradiatively by resonantly promoting a valence electron into the conduction band. We formulate this process as a Brillouin-zone sum over vertical interband transitions weighted by local Th-centered hyperfine matrix elements, which are evaluated directly from all-electron full-potential linearized augmented-plane-wave Bloch spinors. A finite nuclear magnetization model is included to regularize the short-range hyperfine interaction and to account for the Bohr-Weisskopf effect.  After applying scissor shifts to span the experimentally reported \ce{ThO2} band gaps, we find calculated internal-conversion lifetimes in the range of $1-16~\mu\mathrm{s}$. The lifetime increases strongly as band gap approaches $\omega_\mathrm{nuc}$  because the resonant interband phase space at the nuclear transition energy is reduced. For the larger reported \ce{ThO2} gaps, the calculated lifetime is comparable to the measured conversion-electron M\"ossbauer lifetime [Nature {\bf 648}, 300 (2025)]. Our analysis implies that choosing solid-state hosts with bandgap values slightly lower than $\omega_\mathrm{nuc}$ can optimize solid-state nuclear clock performance with internal conversion electron readout.
\end{abstract}
\maketitle

\section{Introduction}
\label{Sec:Intro}

Internal conversion (IC) is one of the oldest known manifestations of direct coupling between nuclear and electronic degrees of freedom. In this process, an excited nucleus decays by resonantly transferring its energy to an electron, which is emitted instead of a radiative photon. The idea originated in early studies of discrete electron lines in radioactive decay, when \citet{Meitner1922} recognized that some electrons accompanying nuclear radiation were secondary atomic electrons rather than primary beta particles. Subsequent theoretical work established that conversion electrons are produced by the same electromagnetic nuclear transition operator responsible for gamma emission, with the transition energy transferred directly to an electron.

Internal conversion soon became an essential tool of nuclear spectroscopy~\cite{RamanHowGoodAre2002}. Its key quantity is the IC coefficient, $\alpha_\mathrm{IC}=\Gamma_{\rm IC}/\Gamma_\gamma$, defined as the ratio of the IC and radiative decay rates. Relativistic calculations of IC coefficients for atoms were developed soon after the quantum theory of nuclear electromagnetic transitions, with Rose's~\cite{Rose1951} K-shell tables providing an early standard reference. Later self-consistent relativistic calculations~\cite{HagerSeltzer1968,BandTrzhaskovskaya1978} made IC coefficients routine inputs to nuclear data evaluation.

The low-energy isomer of \thor{}, with $\hbar \omega_\mathrm{nuc}\simeq 8.35\,\mathrm{eV}$, changes the usual hierarchy of energy scales and opens a range of qualitatively new IC-related phenomena. For ordinary nuclear transitions in the keV--MeV range, the chemical or condensed-matter environment typically enters only as a correction to an essentially atomic IC process. For \thor{}, by contrast, the isomer energy is comparable to valence-electron excitation energies, band gaps, defect-level energies, and work functions. Whether IC is allowed is determined not only by atomic binding energies, but by the broader electronic environment of the thorium nucleus. This point was already emphasized in early theoretical work on the \thor{} isomer, which treated internal conversion, radiative decay, and electronic-bridge processes in neutral Th as competing nuclear-electronic decay channels~\cite{StrizhovTkalya1991}.

In the past few years, direct laser excitation of the isomer has been demonstrated in several solid-state hosts, including \thor{}-doped \caf{} and \lisaf{} crystals~\cite{Tiedau2024Laser,Elwell2024PRL}, thin films of ThF$_4$~\cite{Zhang2024ThF4}, and, most recently, \ce{ThO2} films~\cite{Elwell2025CEMS}. 

Understanding how nuclear degrees of freedom are affected by the solid-state environment is essential for the development of compact, high-performance solid-state nuclear clocks, as well as related searches for ultralight dark matter and fundamental constants variation.
This sensitivity is both a challenge and an opportunity for clock development: near-surface contamination, defects, adsorbates, strain, and altered stoichiometry can modify the local band structure and symmetry at the thorium site, thereby changing the IC rate and potentially shifting the clock frequency. 
These must be managed for optimum clock operation. 
Conversely, by controlling the electronic environment it may be possible to suppress the internal conversion rate.
Given that current crystal-based nuclear clocks operate on transitions with linewidth $\gtrsim 100$~kHz~\cite{Morawetz2026CwAbsorption, ToscaniDeCol2026FeedbackLoop,HuangNuclearClockBased2026}, with a path towards $\sim 10$~kHz linewidth~\cite{Rellergert2010}, if the IC decay lifetime can be increased to $\gtrsim 100~\mu$s an IC-based nuclear clock  may outperform crystal implementation — especially since \ce{ThO2} is not limited by the magnetic dipole broadening that comes from the \ce{^19F} in the Th-doped crystals. 
Such a system has many other practical advantages as the clock platform is decidedly simpler and more robust than crystal implementations, and nuclear state preparation via optical pumping is not limited by the long radiative lifetime.

{A solid-state environment requires a qualitatively different formulation of IC.}
Early work on low-energy IC in \thor{} followed the isolated-atom paradigm~\cite{StrizhovTkalya1991}, whereas in a crystal the relevant electronic states are extended Bloch states organized into bands and, for the IC matrix element, by their local projection onto the thorium site~\cite{Morgan2025ICSolidState,Elwell2025CEMS}. The availability of final states is therefore governed by the band structure, Fermi level, band gap, defect spectrum, and hybridization between Th-centered orbitals and the host material. {Thus, a realistic theory of low-energy IC in solids must combine nuclear spectroscopy with concepts from condensed-matter physics and materials chemistry.}
{In this setting, IC is best viewed not as a small correction to an atomic process, but as a nuclear decay channel embedded in a many-body electronic medium.}

This shift from isolated atoms to embedded nuclei makes it useful to classify \thor{} IC channels in explicitly solid-state terms. In the atomic literature, IC is commonly separated into bound-to-continuum and bound-to-bound processes, depending on whether a bound electron is emitted into the continuum or promoted to another bound state. In a crystal, the corresponding final states are instead organized by the host electronic structure. Following Ref.~\cite{Derevianko2026ColloquiumNuclearClocks}, we distinguish three limiting channels: band-to-defect, defect-to-band, and band-to-band IC.

\begin{itemize}
    \item \emph{Band-to-defect IC---} In \thor{}-doped high-band-gap insulators with $E_\mathrm{gap} > \hbar\omega_\mathrm{nuc}$, such as the hosts used in recent laser-excitation experiments~\cite{Tiedau2024Laser,Elwell2024PRL}, a Th-related excited electronic state may appear as a localized in-gap level at energy $\varepsilon_d < E_\mathrm{gap}$ above the valence-band maximum. If $\varepsilon_d < \hbar\omega_\mathrm{nuc}$, the isomer can decay by promoting a valence-band electron into this defect level~\cite{Morgan2025ICSolidState}.

    \item \emph{Defect-to-band IC---} If the defect level is occupied and $E_\mathrm{gap}-\varepsilon_d < \hbar\omega_\mathrm{nuc}$, the nuclear energy can instead promote the defect electron into the conduction band. This channel provides a mechanism for quenching through photoinduced or defect-assisted electronic dynamics in the host~\cite{Terhune2025Photoinduced,Guan2025-Xray-quenching-Th-CaF2}.

    \item \emph{Band-to-band IC---} In stoichiometric low-band-gap insulators with $E_\mathrm{gap} < \hbar\omega_\mathrm{nuc}$, such as \ce{ThO2}, no localized defect level is required: the isomer can decay by promoting a valence-band electron directly into the conduction band~\cite{Elwell2025CEMS}.
\end{itemize}

The essential point is that, in \thor{} crystals, IC is no longer described by a universal tabulated atomic coefficient. It becomes a host-dependent decay channel of an embedded nucleus, controlled by the local electronic structure near the thorium site. For example, the band-to-defect IC rate derived in Ref.~\cite{Morgan2025ICSolidState} depends explicitly on the valence-band density of states, illustrating how quantities familiar from materials science enter directly into the nuclear decay rate.

\begin{figure}[h!]
{
\centering
\includegraphics[width=\linewidth]{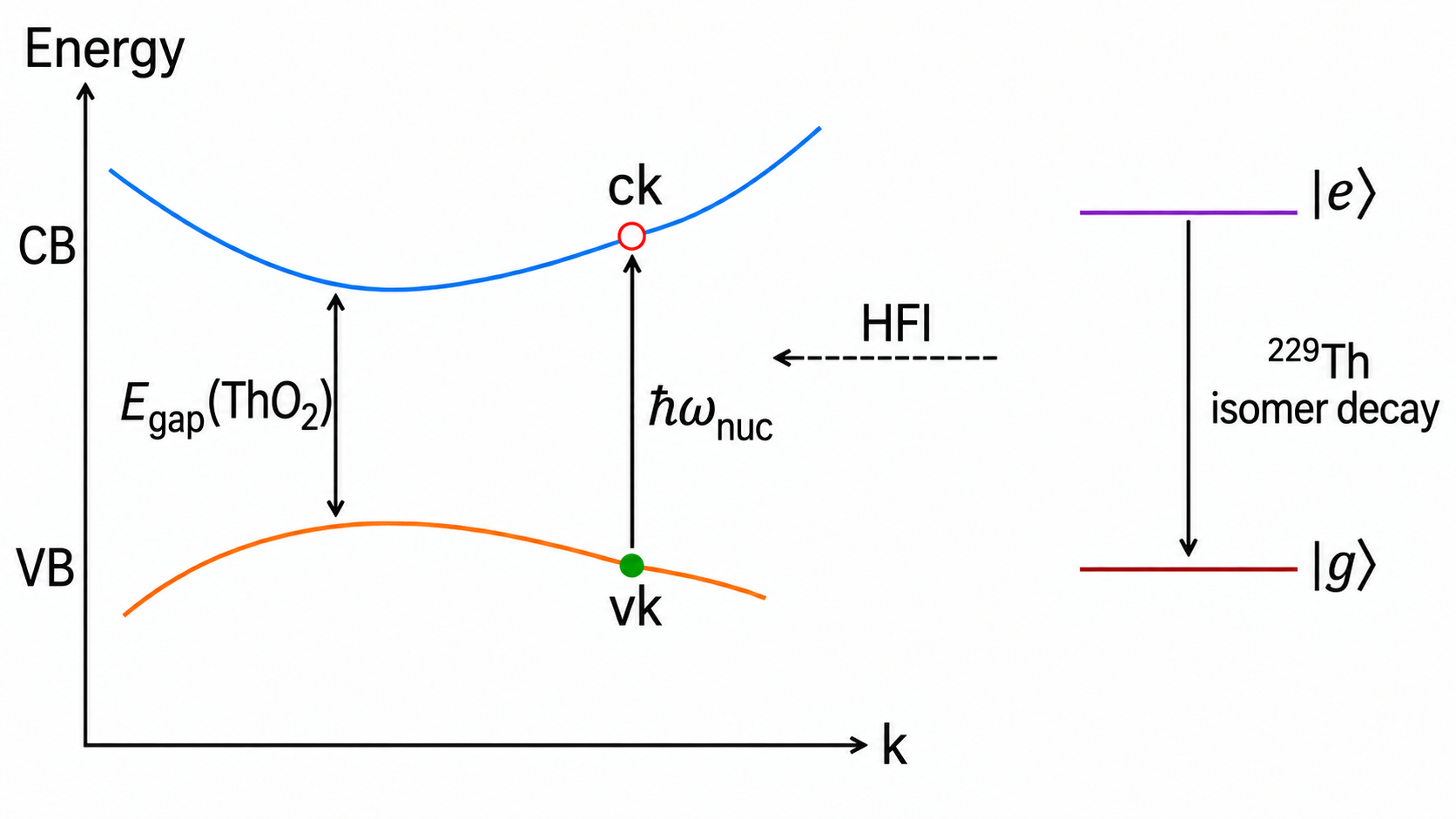}
}
    \caption{
    Schematic of band-to-band internal conversion process in
    \textsuperscript{229}ThO\textsubscript{2}. The quantity
    \(E_{\mathrm{gap}}\)(ThO\textsubscript{2}) denotes the electronic
    band gap of ThO\textsubscript{2}. Since
    \(E_{\mathrm{gap}} < \hbar\omega_{\mathrm{nuc}}\),
    the \textsuperscript{229}Th isomer can decay from \(|e\rangle\)
    to \(|g\rangle\) by resonantly transferring the nuclear transition energy
    \(\hbar\omega_{\mathrm{nuc}}\) through the hyperfine interaction (HFI)
    and promoting an electron from a valence-band (VB) state \(v\mathbf{k}\)
    to a conduction-band (CB) state \(c\mathbf{k}\).
    }
    \label{fig:band_to_band_IC_schematic}
\end{figure}

{In this work we focus on the band-to-band IC channel, which is  relevant for low-band-gap \thor{} compounds.} The band-to-band process considered in our paper is illustrated  in Fig.~\ref{fig:band_to_band_IC_schematic}.
Ref.~\cite{Elwell2025CEMS} demonstrated that, in a low-band-gap host, band-to-band IC can dominate over nuclear radiative decay and provide a fast electronic readout channel enabling a novel platform for a solid-state nuclear clock.  The IC lifetime has been measured directly by laser-based conversion-electron M\"ossbauer spectroscopy~\cite{Elwell2025CEMS}, motivating our {\em ab initio} calculations. {More broadly, the computational framework developed here is not specific to \ce{ThO2}; it can be applied to other stoichiometric \thor{} compounds in which band-to-band IC is energetically allowed.}

The band-to-band IC formalism was developed in Ref.~\cite{Elwell2025CEMS}. In that treatment, the nuclear isomer decays by creating an interband particle-hole pair with total energy $\hbar\omega_\mathrm{nuc}$; equivalently, a valence-band Bloch state is promoted to a conduction-band Bloch state. The IC rate is obtained from Fermi's golden rule, with the hyperfine interaction providing the nuclear-electronic coupling and lattice periodicity enforcing crystal-momentum conservation. The resulting expression separates a local hyperfine matrix element at the Th site from the valence-to-conduction joint density of states at $\hbar\omega_\mathrm{nuc}$. {We review the main steps of this derivation in Sec.~\ref{Formula} to make the notation and implementation self-contained.}

In Ref.~\cite{Elwell2025CEMS}, we also presented a semi-empirical estimate of the IC rate in \ce{ThO2} based on Th-projected densities of states and hyperfine-structure constants of the \ce{^{229}Th^{3+}} ion. That approximation captured the basic scale of the band-to-band IC rate, but it replaced the local crystal-state matrix elements by projected densities of states and atomic hyperfine constants. 

Here, we carry out the calculations using the all-electron full-potential linearized augmented-plane-wave plus local-orbital (FP-LAPW+lo) framework implemented in the Elk code~\cite{ElkCode,ElkManual}. This framework is well suited to the IC problem because it retains the full radial and angular structure of the Bloch spinors inside the near-nucleus region where the hyperfine interaction is localized. This local information is essential, since the IC rate depends not only on the energetic availability of interband transitions, but also on the electronic states at the $^{229}$Th site~\cite{Morgan2025ICSolidState,Elwell2025CEMS}. Our implementation therefore evaluates the field-weighted interband matrix elements directly, rather than inferring them from projected densities of states combined with atomic hyperfine constants. In this sense, the FP-LAPW+lo framework provides a natural platform for an \textit{ab initio} treatment of host-dependent internal conversion in solid-state $^{229}$Th systems.

This paper is organized as follows. In Sec.~\ref{Formula} we review the formalism for the band-to-band IC rate in a crystal and express the rate in terms of nuclear and electronic matrix elements. In Sec.~\ref{Elk_modification} we describe our FP-LAPW+lo-based implementation, including the evaluation of the local Th-centered hyperfine matrix elements from all-electron Bloch spinors and the finite nuclear-magnetization model. The main results for \ce{ThO2}, including the electronic structure, band-gap correction, and calculated IC rates and lifetimes, are presented in Sec.~\ref{result}. In Sec.~\ref{result} we discuss the limitations of the present treatment and present insights into the behavior of our computed IC rate as a function of band gap. Finally, conclusions are drawn in Sec.~\ref{Sec:Conclusion}. Additional numerical details and angular reduction of various matrix elements are collected in the appendices. 

Unless specified otherwise, we use atomic units ($|e|=\hbar=m_e \equiv 1$) and Gaussian units for electromagnetism. In these units, the nuclear magneton is $\mu_N=\alpha/(2m_p)$, where proton mass $m_p$ is expressed in units of electron mass $m_e$, and $\alpha$ is the fine-structure constant. Elementary charge $e>0$, so that the electron charge is $-e$.

\section{Formalism}
\label{Formula}

In this section, we review the band-to-band IC-rate derivation from our earlier work~\cite{Elwell2025CEMS} and recast the result in a form suitable for numerical evaluation on a discretized reciprocal space mesh.  

In the IC process, the nucleus resonantly transfers its excitation to the crystal.
In the exact many-body description, the nuclear transition couples through the
hyperfine interaction (HFI) to the full electronic and phononic excitation spectrum of the solid.
Here, we adopt an independent-particle approximation, in which
the nuclear energy is transferred by promoting a valence-band electron into the conduction band. The accuracy of this approximation  is discussed in Sec.~\ref{result}.
Because the HFI is cell-periodic in the crystal, its matrix element between
Bloch states conserves crystal momentum, so the direct electronic excitations are
vertical transitions at fixed crystal momentum $\mathbf{k}$. 

The IC decay in $^{229}$Th connects 
the $3/2^+[631]$ isomeric state $\ket{e}$ to the 
$5/2^+[633]$ nuclear ground state $\ket{g}$. In the Nilsson description, these are two nearby same-parity bandheads in a deformed odd-neutron nucleus. Since the two states have the same parity, the lowest-multipolarity  IC decay channels are M1 and E2.
Although E2 contributions can be comparable to or larger than M1 contributions in some internal-conversion and electronic-bridge channels~\cite{Bilous2018,Nickerson2021}, the ideal Th site in fluorite \ce{ThO2} has cubic $O_h$ symmetry, for which the E2 contribution vanishes by symmetry. Then the IC process is mediated by the M1 hyperfine interaction  ${H}_{\mathrm{HFI}}^{(M1)}$.

The HFI operator  can be represented as a rotationally invariant product of the nuclear transition magnetic moment operator $\boldsymbol \mu$ and the electronic tensor $\mathbf T^{(e)}$,
\begin{equation}
 H_{\mathrm{HFI}}^{(M1)}
=
{\boldsymbol{\mu}}\cdot
{\mathbf T^{(e)}} = \sum_{q=-1}^{+1} (-1)^q \mu_{-q} T^{(e)}_q\,,
\label{eq:Hhfi_muDotT}
\end{equation}
where we expanded the scalar product in terms of spherical components of rank-1 tensors.
In the present work, considering that Elk is not a fully-relativistic Dirac code, we will use the non-relativistic limit of the electronic rank-1 tensor ${\mathbf T^{(e)}}$, as described in Sec.~\ref{Elk_modification}. 

We denote the full two-component Bloch spinor by
\(\Psi_{n\mathbf k}(\mathbf r)\equiv \braket{\mathbf r}{n\mathbf k}\),
where \(n\) labels a Kohn--Sham spinor band and \(\mathbf k\) is the
crystal momentum. The spinor satisfies the Bloch condition
\(\Psi_{n\mathbf k}(\mathbf r+\mathbf T)
=
e^{i\mathbf k\cdot\mathbf T}\Psi_{n\mathbf k}(\mathbf r)\)
for any lattice vector \(\mathbf T\). Therefore, by Bloch's theorem, it
can be written as
\(\Psi_{n\mathbf k}(\mathbf r)
=
e^{i\mathbf k\cdot\mathbf r}u_{n\mathbf k}(\mathbf r)\), where
\(u_{n\mathbf k}(\mathbf r)\) is a cell-periodic two-component spinor.

With this convention, the Bloch spinors are normalized over the primitive
unit cell as
\begin{equation}
\int_{\Omega} d^3r\,
\Psi_{n\mathbf k}^{\dagger}(\mathbf r)
\Psi_{n'\mathbf k}(\mathbf r)
= \delta_{nn'} ,
\label{eq:Bloch_normalization}
\end{equation}
where \(\Omega\) is the primitive unit-cell volume. This normalization differs
from that used in Ref.~\cite{Elwell2025CEMS} and in addition 
includes the possibility of including spin-orbit coupling (SOC) by absorbing electron spin degrees of freedom into the spinor parameterization of wave-function. Notice that in the spin-independent scalar-relativistic problem, each spatial band is twofold spin degenerate. After SOC is included, these states are more naturally labeled as spinor bands; the number of one-electron eigenstates is correspondingly doubled, although crystal symmetries may keep them pairwise degenerate.

For a periodic crystal, translational symmetry requires conservation of crystal momentum. Thus, the matrix element between an initial state with the nucleus in $|e\rangle$ and an electron in a
valence-band Bloch state $|v\mathbf{q}\rangle$, and a final state with the nucleus in $|g\rangle$ and an electron in a conduction-band Bloch state
$|c\mathbf{k}\rangle$ reduces to
\begin{equation}
\langle g; c\mathbf{k} |
{H}_{\mathrm{HFI}}^{(M1)}
| e; v\mathbf{q}\rangle
=
\delta(\mathbf{k}-\mathbf{q})\,
W^{ge}_{cv}(\mathbf{k}),
\label{eq:W_definition}
\end{equation}
where $\ket{e} \equiv \ket{I_e=3/2,M_e}$ and $\ket{g} \equiv \ket{I_g=5/2,M_g}$ denote the nuclear isomeric and ground states with their magnetic quantum numbers,
$v$ and $c$ label valence and conduction bands, and $\mathbf{q}$ and $\mathbf{k}$
are crystal momenta. The delta function enforces crystal-momentum conservation. The matrix element \(W^{ge}_{cv}(\mathbf k)\) couples
the nuclear transition \(\ket{e}\to\ket{g}\) to the electronic excitation
\(\ket{c\mathbf k}\leftarrow\ket{v\mathbf k}\) at fixed \(\mathbf k\),
\begin{equation}
W^{ge}_{cv}(\mathbf k)
\equiv
\int_{\Omega} d^3r\,
\Psi_{c\mathbf k}^{\dagger}(\mathbf r)\,
\mel{g}{H_{\mathrm{HFI}}^{(M1)}}{e}\,
\Psi_{v\mathbf k}(\mathbf r).
\label{Eq:Wmel}
\end{equation}

The band-to-band IC rate is given by Fermi's golden
rule~\cite{Elwell2025CEMS},
%\begin{equation}
\begin{align}
\Gamma_{\mathrm{IC}}
&=
\frac{2\pi}{\hbar}
\frac{\Omega}{(2\pi)^3}
\frac{1}{[I_e]}
\sum_{M_g,M_e}
\sum_{c,v}
\int_{\mathrm{BZ}} d^3k\,
\left|
W^{ge}_{cv}(\mathbf k)
\right|^2
\nonumber\\
&\quad \times
\delta\left(
\varepsilon_{c\mathbf k}
-
\varepsilon_{v\mathbf k}
-
\hbar\omega_{\mathrm{nuc}}
\right).
\label{eq:rate_volume}
\end{align}  
%\end{equation}
Here, BZ denotes the first Brillouin zone, and we use the shorthand $[I_e]\equiv 2I_e+1$. The factor $1/(2I_e+1)$ averages over the initial nuclear magnetic sublevels $M_e$, while the sum over $M_g$ runs over the final ground-state nuclear sublevels. 

The quantities $\varepsilon_{v\mathbf{k}}$ and $\varepsilon_{c\mathbf{k}}$ are the valence- and conduction-band energies at crystal momentum $\mathbf{k}$, respectively. The Dirac delta function enforces energy conservation for an interband transition driven by the nuclear transition energy $\hbar\omega_{\mathrm{nuc}}$.

In numerical evaluation, the k-space is discretized and we 
convert a Brillouin-zone integral in Eq.~\eqref{eq:rate_volume} into the sum, $\frac{\Omega}{(2\pi)^3} \int_{\mathrm{BZ}} d^3k \cdots\, \rightarrow \sum_\mathbf{k} w_\mathbf{k}  \cdots $, where \(w_{\mathbf k}\) is the numerical integration weight associated with
the \(k\)-point. The discretized rate is therefore
\begin{align}
\Gamma_{\mathrm{IC}}
&=
\frac{2\pi}{\hbar}
\frac{1}{[I_e]}
\sum_{M_g,M_e}
\sum_{c,v}
\sum_{\mathbf k} w_{\mathbf k}
\left|
W^{ge}_{cv}(\mathbf k)
\right|^2
\nonumber\\
&\quad \times
\delta\left(
\varepsilon_{c\mathbf k}
-
\varepsilon_{v\mathbf k}
-
\hbar\omega_{\mathrm{nuc}}
\right).
\label{eq:rate_discrete}
\end{align}  

With the spherical component decomposition~\eqref{eq:Hhfi_muDotT} of the HFI, application of the Wigner-Eckart theorem for the nuclear M1 matrix element, and summation over nuclear magnetic quantum numbers,
\begin{equation}
\Gamma_{\mathrm{IC}}
=
\frac{2\pi}{\hbar}
\frac{1}{2I_e+1}
\left|
\left\langle
I_g
\left\|
\mu
\right\|
I_e
\right\rangle
\right|^2
S_{\mathrm{el}}(\omega_{\mathrm{nuc}}).
\label{eq:GammaIC_in_terms_of_Sel}
\end{equation}
Here the electronic spectral factor reads
\begin{equation}
\begin{aligned}
S_{\mathrm{el}}(\omega_{\mathrm{nuc}})
&=
\frac{1}{3}
\sum_{\mathbf{k}} w_{\mathbf{k}}
\sum_{v,c}
\sum_{q=-1}^{1}
\left|
I_q(c,v,\mathbf{k})
\right|^2 \times\\
& \quad \quad \delta(
\varepsilon_{c\mathbf{k}}
-\varepsilon_{v\mathbf{k}}
-\hbar\omega_{\mathrm{nuc}}) \,,
\end{aligned}
\label{eq:Sel_def}
\end{equation}
where the electronic HFI matrix elements are ($q=-1,0,+1$)
\begin{equation}
I_q(c,v,\mathbf k)
=
\int_{\Omega} d^3r\,
\Psi_{c\mathbf k}^{\dagger}(\mathbf r)\,
T_q^{(e)}(\mathbf r)\,
\Psi_{v\mathbf k}(\mathbf r).
\label{eq:Iq_braket}
\end{equation}

Further, we use the $I_e\rightarrow I_g$ reduced M1 transition
probabilities common in nuclear physics
\begin{equation}
B(M1)
=
\frac{3}{4\pi}
\frac{1}{2I_e+1}
\left|
\left\langle
I_g
\left\|
\mu
\right\|
I_e
\right\rangle
\right|^2 .
\label{eq:BM1_mu_relation}
\end{equation}
Notice that the magnetic-moment operator $\boldsymbol \mu$ used in our derivations
differs from the nuclear-physics convention for M1 radiative transition operator,
%${M}_{1\mu},
%=\sqrt{4\pi/3}\,{M}_{1}$.
$\boldsymbol{M}^{(n)}_{1}
=\sqrt{{3}/{4\pi}}\, \boldsymbol \mu$; the $3/4\pi$ pre-factor in Eq.~\eqref{eq:BM1_mu_relation} accounts for this difference.
With $B(M1)$, the IC rate can also be expressed as
\begin{equation}
\Gamma_{\mathrm{IC}}
=
\frac{2\pi}{\hbar}
\frac{4\pi}{3}
B(M1)
S_{\mathrm{el}}(\omega_{\mathrm{nuc}}),
\label{eq:GammaIC_in_terms_of_Sel_BM1}
\end{equation}
explicitly factorizing the nuclear, $B(M1)$, and electronic $S_{\mathrm{el}}(\omega_{\mathrm{nuc}})$ quantities. In Sec.~\ref{Finite_nuclear}, we introduce corrections to this factorized form that arise from the penetration of electronic wave functions into the nuclear volume.

Finally, the internal-conversion lifetime is an inverse of the IC rate, \(\tau_{\mathrm{IC}}=1/\Gamma_{\mathrm{IC}}\).
The evaluation of the electronic matrix elements in the FP-LAPW+lo method is described in the following sections.

\section{FP-LAPW+lo implementation}
\label{Elk_modification}

The IC rate~\eqref{eq:GammaIC_in_terms_of_Sel_BM1} is factorized into the nuclear and the electronic contribution. We use an all-electron, full-potential LAPW framework to evaluate the electronic part.
In the  FP-LAPW+lo method~\cite{Andersen1975,Wimmer1981}, the localized \thor{}-centered HFI matrix elements can be computed directly from Bloch spinors. 
In the LAPW method, the unit cell is divided into two regions:
non-overlapping atomic spheres, called muffin-tin (MT) regions, and the
remaining interstitial region (IR). The MT sphere around Th acts as an
atom-centered inner region, where the electronic wave function is
resolved into radial functions and spherical harmonics about the nucleus. The IR is described by plane waves, which efficiently represent
the more delocalized bonding and band character of the crystal wave
functions. The full Bloch spinor is continuous across the MT boundary,
so the two descriptions together represent a single crystal eigenstate.
We evaluate the hyperfine matrix element over the MT region because
the M1 hyperfine operator is strongly localized near the Th nucleus,
making the IR contribution negligible for this local operator.

For each second-variational\footnote{In the second-variational treatment,
spin-orbit coupling is included by diagonalizing the spin-dependent Hamiltonian
in a basis of previously obtained scalar-relativistic LAPW eigenstates.} eigenstate, the cell-periodic Bloch spinor \(\psi_{n\mathbf{k}}\) associated with band \(n\) and crystal momentum \(\mathbf{k}\) is expanded inside the Th muffin-tin sphere of radius $R_\mathrm{MT}^\mathrm{Th} \approx 2.56 \, \mathrm{bohr}$ 
as
\begin{equation}
\psi_{n\mathbf{k}}(\mathbf r)
=
\sum_{lm\sigma}
R^{n\mathbf{k}}_{lm\sigma}(r)\,
Y_{lm}(\hat{\mathbf r})\chi_\sigma, \quad r< R_\mathrm{MT}^\mathrm{Th} \,.
\label{Eq:Psi-MT-expansion}
\end{equation}
Here \(R^{n\mathbf{k}}_{\ell m\sigma}(r)\) are the spin-resolved radial amplitudes of the Bloch spinor inside the Th muffin-tin sphere, \(Y_{\ell m}(\hat{\mathbf r})\) are spherical harmonics, and \(\chi_\sigma\) are the conventional spinors. Here and below, \(\mathbf r\) is measured relative to the Th nucleus. 

{Bloch functions~\eqref{Eq:Psi-MT-expansion} are two-component spinors; they are not fully relativistic four-component Dirac bispinors.}
Relativistic effects in Elk are treated using the standard
scalar-relativistic plus second-variational spin–orbit scheme of the
FP-LAPW+lo method. The first-variational LAPW basis is constructed from solutions
of scalar-relativistic radial equations obtained by reducing the Dirac equation
while omitting the explicit spin–orbit interaction. This treatment retains the
spin-independent relativistic kinematics, conventionally associated with the
mass–velocity and Darwin corrections, as well as higher-order scalar-relativistic
contributions. Spin–orbit coupling is subsequently included by diagonalizing
the spin–orbit Hamiltonian in the basis of first-variational eigenstates
\cite{KoellingHarmon1977,SinghNordstrom2006,ElkManual}.
Core states are treated separately as atomic-like states obtained by solving
the radial Dirac equation in the spherical part of the Kohn–Sham potential.

Consistent with the two-component description of the valence and conduction
Bloch states, we use the Pauli-reduced form of the electronic M1 HFI tensor $\mathbf T^{(e)}$. A spherical component of the HFI tensor
in atomic-Gaussian units can be written in the standard Pauli-limit
orbital, spin-dipole, and Fermi-contact decomposition as~\cite{Schwartz1955HFS,Jonsson1993HFS}
\begin{equation}
\begin{aligned}
T^{(e)}_{q}(\mathbf r)
= \alpha \bigg[
& \frac{l_q}{r^3}
+ \frac{g_e}{4r^3}
\left(
3\hat r_q\,{\boldsymbol{\sigma}}\cdot\hat{\mathbf r}
-\sigma_q
\right)  \\
& + \frac{2\pi g_e}{3}\,
\sigma_q\,\delta(\mathbf r)
\bigg] .
\end{aligned}
\label{Pauli_T1q_point}
\end{equation}
Here $q=-1,0,+1$ labels spherical components.
$\boldsymbol{l}$ is the orbital angular momentum operator, ${\boldsymbol{\sigma}}$ is the vector of Pauli matrices, $\hat{\mathbf r}$ is the unit vector and $r=|\mathbf r|$ is the electron-nucleus distance. $g_e = 2.0023\ldots$ is the electron $g$-factor, $\alpha$ is the fine-structure constant.  This approximation retains the leading orbital, spin-dipole, and
Fermi-contact magnetic couplings but misses the higher-order $(\alpha Z)^2$ contributions. This expression is for a point nuclear magnetic dipole; we will generalize it to include finite nuclear size below.

The non-relativistic limit~\eqref{Pauli_T1q_point} of the ${\mathbf T^{(e)}}$ operator is conventionally  decomposed into the orbital,
spin-dipole, and Fermi-contact contributions,
\begin{equation}
T^{(e)}_{q}(\mathbf r)
=
 T^{(\mathrm{orb})}_{q}(\mathbf r)
+
T^{(\mathrm{sd})}_{q}(\mathbf r)
+
 T^{(\mathrm{FC})}_{q}(\mathbf r),
\label{eq:T_decomp_append}
\end{equation}
where \( T^{(\mathrm{orb})}_{q}\), \( T^{(\mathrm{sd})}_{q}\), and
\( T^{(\mathrm{FC})}_{q}\) stand for the orbital, spin-dipole, and
Fermi-contact terms, respectively. 
The orbital term describes the coupling
of the nuclear magnetic dipole to the magnetic field generated by the
electronic orbital motion. The spin-dipole term describes the anisotropic
dipole--dipole coupling to the electronic spin density. The Fermi-contact term gives the isotropic contribution associated with the electronic spin density at the nucleus. For a point nuclear magnetic dipole, these contributions have singular short-range structure: the orbital and spin-dipole terms contain $r^{-3}$ singularities, whereas the contact term is localized at $r=0$. For an extended nuclear magnetization distribution, this short-range structure is regularized.

Considering the short range nature of $T^{(e)}_{q}(\mathbf r)$, the  electronic matrix element~\eqref{eq:Iq_braket} for a valence-to-conduction transition at
fixed crystal momentum \(\mathbf{k}\) is evaluated as 
\begin{equation}
I_q(c,v,\mathbf k)
=
\int_{\Omega_{\mathrm{MT}}^{\mathrm{Th}}}
d^3r\,
\psi_{c\mathbf{k}}^\dagger(\mathbf r)\,
T^{(e)}_{q}(\mathbf r)\,
\psi_{v\mathbf{k}}(\mathbf r),  
\label{Eq:I-integrals}
\end{equation}
where \(\Omega_{\mathrm{MT}}^{\mathrm{Th}}\) is the Th muffin-tin
region. We evaluate
these matrix elements directly from the all-electron crystal wave functions inside the \thor{} muffin-tin sphere. 
To this end, we exploit the expansion of MT wave functions in
spherical harmonics, Eq.~\eqref{Eq:Psi-MT-expansion}, to simplify the
electronic integrals in Eq.~\eqref{Eq:I-integrals}. The resulting angular
reduction is presented in Appendix~\ref{app:angular_reduction}. There,
the angular integrations are carried out analytically, which accelerates the calculations and improves numerical accuracy.

\subsection{Bohr-Weisskopf effect for transition M1 moment}
\label{Finite_nuclear}

The magnetic-dipole HFI is commonly evaluated in the
point-dipole approximation, where the nuclear magnetic moment is localized at
the origin. For heavy nuclei this treatment is insufficient since relativistic
$s_{1/2}$ and $p_{1/2}$ wave functions penetrate the nuclear volume, making the
short-range hyperfine matrix element potentially sensitive to the finite magnetization
distribution. This resulting correction is called the Bohr--Weisskopf (BW) effect
\cite{BohrWeisskopf1950,Stroke2000,Roberts2022}. The original BW effect has been considered for expectation values of nuclear magnetic moment. Here we generalize it to transition moments.

In the present IC rate calculation, the BW correction is included through
a finite nuclear magnetization model for the local magnetic hyperfine operator.
We use a uniformly magnetized spherical nucleus of radius $R_N$, described by
the normalized magnetization density
\begin{equation}
f_m(\mathbf r)
=
\frac{1}{V_N}\Theta(R_N-r),
\qquad
\int d^3r\,f_m(\mathbf r)=1 .
\label{eq:uniform_magnetization_density}
\end{equation}
Here \(V_N=4\pi R_N^3/3\) is the nuclear volume and \(\Theta(x)\) is the
Heaviside step function. The value of \(R_N=5.733~\mathrm{fm}\) is the default Elk finite-nuclear-size
radius for Th, obtained from an empirical \(A(Z)\) fit~\cite{Andrae2002}
and the charge-radius parametrization~\cite{Angeli2004}.
Outside the nucleus, $r>R_N$, the
magnetic field is identical to that of a point magnetic dipole, whereas inside
the nucleus the hyperfine operator is regularized.

The atomic IC rate calculations~\cite{Rose1951,HagerSeltzer1968,BandTrzhaskovskaya1978,RamanHowGoodAre2002} involve a closely related short-distance sensitivity: relativistic bound and continuum electron wave functions penetrate the nuclear volume, so the conversion matrix element can depend on the assumed nuclear transition charge or current distribution inside the nucleus. This dependence is commonly treated as a nuclear-penetration correction, with tabulations using model prescriptions for the nuclear region together with relativistic atomic electron wave functions. In the present solid-state study, the sensitivity enters through the local M1 hyperfine matrix element between Bloch states at the Th site. We therefore use the uniformly magnetized sphere as a simple regularization of the M1 operator inside \(R_N\).

For a uniformly magnetized sphere,
the orbital, spin-dipole, and Fermi-contact operators in Eq.~\eqref{eq:T_decomp_append} are
\begin{align}
T^{(\mathrm{orb})}_{q}(\mathbf r)
&=
\alpha\,\frac{{l}_q}{\max(r,R_N)^3},
\\[6pt]
T^{(\mathrm{sd})}_{q}(\mathbf r)
&=
\alpha\,\frac{g_e}{4}\,\Theta(r-R_N)\,
\frac{3\hat r_q\bigl({\boldsymbol{\sigma}}\cdot\hat{\mathbf r}\bigr)-{\sigma}_q}{r^3},
\\[6pt]
T^{(\mathrm{FC})}_{q}(\mathbf r)
&=
\alpha\,\frac{2\pi g_e}{3}\,\frac{1}{V_N}\,\Theta(R_N-r)\, {\sigma}_q.
\end{align}
Here  the factors \(\Theta(r-R_N)\) and
\(\Theta(R_N-r)\) restrict the corresponding terms to the exterior and interior
of the nucleus, respectively.
This provides a numerically stable evaluation of integrals. Compared to the point-like magnetic dipole, introduction of the finite-nuclear-magnetization modifies only the radial kernels of the orbital, spin-dipole, and Fermi-contact operators, see Appendix~\ref{app:angular_reduction}; the corresponding angular selection rules in the muffin-tin partial-wave basis are unchanged.

We emphasize that the uniformly magnetized sphere should be regarded only as a toy model for a microscopic description of the nuclear structure of \thor{}. Future progress in the microscopic theory of \thor{} nuclear structure should enable more realistic refinements of this model. 

In the IC rate~\eqref{eq:GammaIC_in_terms_of_Sel_BM1}, the nuclear transition
strength is introduced separately through the reduced transition
probability $B(M1)$; this value is a purely nuclear-structure quantity. The finite-magnetization model enters only through the
short-distance electronic HFI operator and is used to estimate the
BW sensitivity of the local electronic matrix elements.

As a validation of our approach and numerical implementation, we computed the magnetic-dipole hyperfine-structure constant $A$ for the ground $5f_{5/2}$ state of \ce{^{229}Th} ion using Eq.~\eqref{Pauli_T1q_point}, the PBE functional and the Elk code employed in the IC rate estimate.  This constant is determined by the diagonal matrix element of $\mathbf T^{(e)}$. Our numerical value of $A_{5f_{5/2}}=80.2$~MHz is smaller by 2\% than the experimental value~\cite{Zitzer2025LaserSpectroscopyHyperfine} of $82.0(2)$~MHz.

\section{Results and discussion}
\label{result}

In this section, we present the result of our {\em ab initio} computation of the
band-to-band IC rate in a \(^{229}\mathrm{ThO}_2\) crystal. The rate is obtained by combining
the local Th-centered hyperfine matrix elements, evaluated from
all-electron LAPW Bloch spinors, with the resonant interband
transition spectrum at the nuclear transition energy
\(\hbar\omega_{\mathrm{nuc}}\) using Eq.~\eqref{eq:GammaIC_in_terms_of_Sel_BM1}. Unless stated otherwise, the IC rate results
are reported using\footnote{The \(B(M1)\) values are often quoted in Weisskopf units (W.u.); \(1~{\rm W.u.}=1.790\,\mu_N^2\).
}  \(B(M1)=0.0220(6)~\mathrm{W.u.}\)
inferred from the radiative lifetime measurement in a \ce{^{229}Th}:\ce{CaF2} doped crystal~\cite{Tiedau2024Laser}.

\begin{figure*}[t]
\centering
\includegraphics[width=16cm, height=10cm]{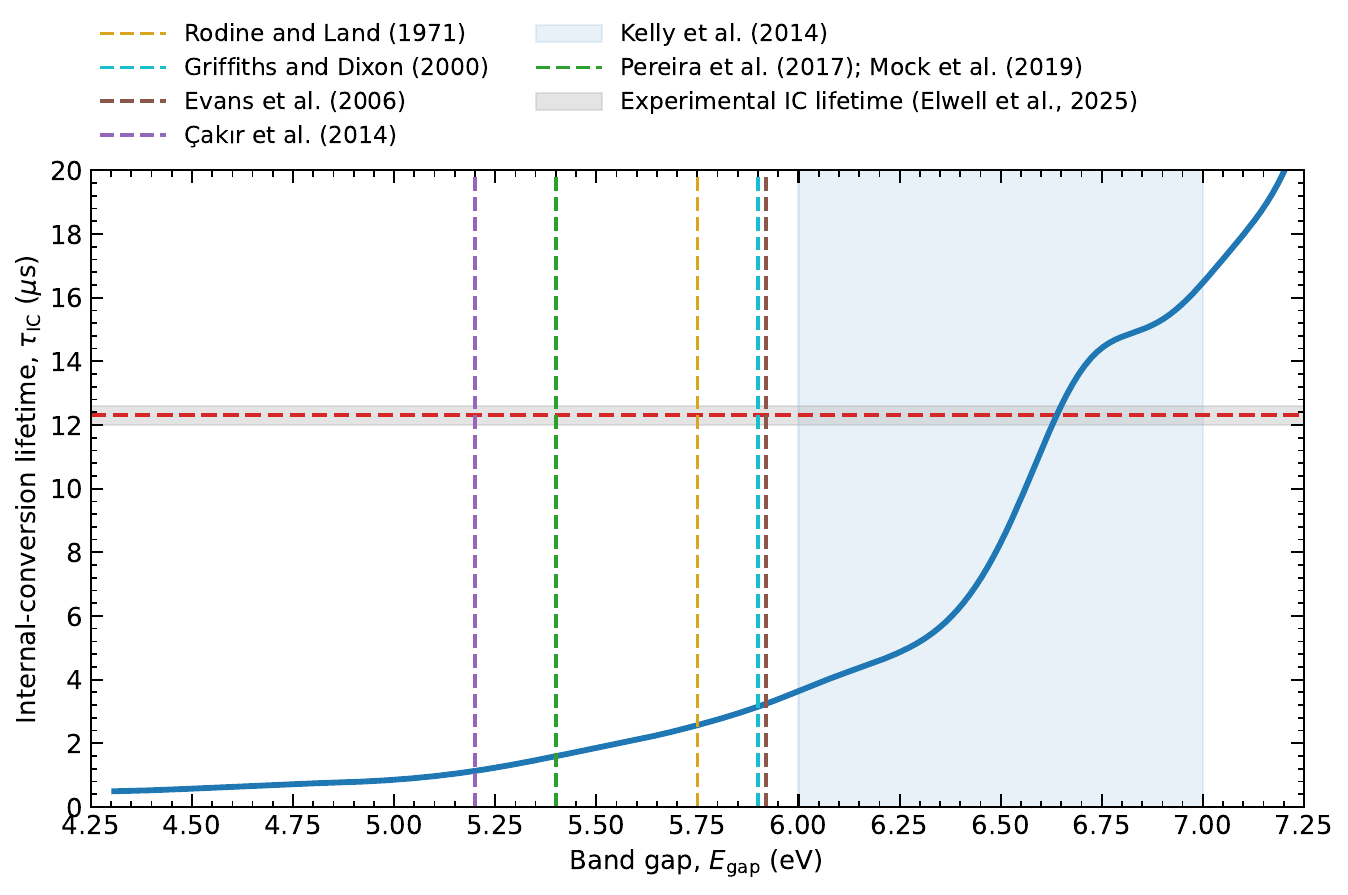}
\caption{
Computed band-to-band internal-conversion lifetime, $\tau_{\mathrm{IC}}$, in
$^{229}\mathrm{ThO}_2$ as a function of the scissor-shifted band gap.
The solid curve shows $\tau_{\mathrm{IC}}$ obtained by rigidly shifting
the conduction bands. The Brillouin zone was sampled using a \(32\times32\times32\) \(k\)-point mesh. The horizontal dashed line indicates
the experimental conversion-electron M\"ossbauer lifetime,
$\tau_{\mathrm{IC}}=12.3(3)~\mu\mathrm{s}$, reported for
$^{229}\mathrm{ThO}_2$ by \citet{Elwell2025CEMS}.
The vertical dashed lines mark the reported experimental electronic gap values from \citet{Rodine1971ThO2},
\citet{Griffiths2000ThO2}, \citet{Evans2006ThO2Gap}, \citet{Cakir2014ThOx}, \citet{Pereira2017ThO2Nanoparticles}, and \citet{Mock2019ThO2BandGap}. The blue shaded region 
shows the $6$--$7~\mathrm{eV}$ spread reported by \citet{Kelly2014ThO2}.
}
\label{fig:scissor_exp_bandgaps}
\end{figure*}

We first assess the convergence
of the underlying electronic structure with respect to
Brillouin-zone sampling. We then use rigid scissor shifts of the
conduction bands to examine the sensitivity of the IC lifetime to
the \(\mathrm{ThO}_2\) band gap. Finally, we analyze the convergence
of the finite-\(k\)-mesh transition sum and discuss the dependence
of the absolute rate on the adopted nuclear \(B(M1)\) value.

Elk input parameters used in our numerical work are listed in Appendix~\ref{Sec:Numerical}. This appendix also provides  convergence tests and studies the sensitivity of IC rate values to 
the smearing width $\sigma$  for the
finite-width representation of the energy-conserving delta function in the IC rate~\eqref{eq:rate_discrete},
\begin{equation}
\begin{aligned}
\delta(x) \rightarrow \delta_\sigma(x) =
\frac{1}{\sigma}
\frac{\exp(-x/\sigma)}
{\left[1+\exp(-x/\sigma)\right]^2},
\end{aligned}
\label{eq:fd_broadening}
\end{equation}
where
$x \equiv
\varepsilon_{c\mathbf{k}}
-
\varepsilon_{v\mathbf{k}}
-
\hbar\omega_{\mathrm{nuc}}$. We use this $\delta$-function representation for the numerical evaluation of the IC rate on a discretized $\mathbf{k}$-mesh. Unless stated otherwise, we employ $\sigma=0.1~\mathrm{eV}$.

The convergence of the ThO$_2$ ground-state energy and Kohn-Sham band gap
with respect to the Brillouin-zone sampling is shown in Table~\ref{tab:bandgap_totalenergy_kmesh} of Appendix~\ref{Sec:Numerical}. We find that over the range from $10\times10\times10$ to $32\times32\times32$ 
$k$-point meshes both the band gap and the
total energy are well converged. The converged PBE band gap is $E_\mathrm{gap}^{\rm PBE}=4.31~{\rm eV}$,
which is smaller than experimental band gaps for ThO$_2$  compiled in
Table~\ref{tab:ThO2_experimental_bandgaps_lifetimes}. The measured band gaps range from $5.2 \, \mathrm{eV}$ to $7 \, \mathrm{eV}$.

\begin{table*}[ht!]
\caption{
The reported ThO$_2$ band gaps from different experimental methods and sample types,
together with the corresponding calculated IC lifetimes. For each, the
conduction-band energies were shifted rigidly corresponding to the PBE Kohn--Sham gap,
$E_\mathrm{gap}^{\rm PBE}=4.31~{\rm eV}$.  The IC lifetimes $\tau_{\rm IC}$ were obtained using a $32\times32\times32$ $k$-point mesh,
with $\sigma=0.10~{\rm eV}$ and $B(M1)=0.022$ W.u.
}
\label{tab:ThO2_experimental_bandgaps_lifetimes}
\begin{ruledtabular}
\begin{tabular}{llcc}
Reference & Measurement and sample & $E_\mathrm{gap}^{\rm exp} ({\rm eV})$  & $\tau_{\rm IC} (\mu{\rm s})$ \\
\hline
\citet{Cakir2014ThOx}
& XPS/UPS+REELS, thin film
& $\sim 5.2$
& 1.13 \\
\citet{Pereira2017ThO2Nanoparticles}
& UV--vis, nanoparticles
& $\sim 5.4$
& 1.59 \\
\citet{Mock2019ThO2BandGap}
& VUV ellipsometry, single crystal
& 5.4
& 1.59 \\
\citet{Rodine1971ThO2}
& optical/TL, single crystal
& 5.75
& 2.57 \\
\citet{Griffiths2000ThO2}
& defect spectroscopy, single crystal
& $\sim 5.9$
& 3.15 \\
\citet{Evans2006ThO2Gap}
& VUV ellipsometry, thin film
& $\sim 5.9$
& 3.15 \\
\citet{Kelly2014ThO2}
& UPS/IPES, single crystal
& $6$--$7$
& $3.64$--$16.47$ \\
\end{tabular}
\end{ruledtabular}
\end{table*}

To align the calculated PBE band gap with
experimental values, we apply a rigid scissor shift  to the
conduction-band orbital energies. This shift is not intended to represent a full
quasiparticle calculation. Rather, it provides a controlled way to examine how
the IC rate depends on the electronic transition spectrum near the fixed nuclear
transition energy $\hbar\omega_{\rm nuc}$. Table~\ref{tab:icrate_scissors_bandgaps_sigma010} shows the convergence of the IC rate and lifetime with respect to the $k$-mesh size for
two representative band gaps. Both rates vary by a few per cent depending on the $k$-mesh size.

Table~\ref{tab:ThO2_experimental_bandgaps_lifetimes} lists the reported ThO$_2$ band gaps and the corresponding computed IC lifetimes $\tau_{\rm IC}$
obtained by applying the scissor shift to the PBE Kohn-Sham band gap. It shows that the spread
in reported band gaps translates into a substantial spread in the computed $\tau_{\rm IC}$. 
This dependence of $\tau_{\rm IC}$ on $E_\mathrm{gap}$ is further illustrated by Fig.~\ref{fig:scissor_exp_bandgaps}. The IC lifetime increases
as the band gap is shifted upward. Qualitatively, as the band-gap increases there are fewer particle-hole pairs available for excitations satisfying the $\varepsilon_{c\mathbf k}-\varepsilon_{v\mathbf k}= \hbar\omega_{\mathrm{nuc}}$ energy conservation, reducing the rate and thereby increasing $\tau_{\rm IC}$. For  gaps in the $5.2$--$5.9~{\rm eV}$ range, the computed lifetimes are a few microseconds, shorter than the measured conversion-electron M\"ossbauer lifetime of
$12.3(3)~\mu{\rm s}$~\cite{Elwell2025CEMS}. In contrast, when the gap is within the $6$--$7~{\rm eV}$ range of Ref.~\cite{Kelly2014ThO2}, the calculated lifetime is comparable to 
the experimental $\tau_{\rm IC}$. This indicates that the band-to-band IC rate in ThO$_2$ is highly
sensitive to the electronic excitation spectrum near $\hbar\omega_{\rm nuc}$, and that quantitative comparison with experiment requires an accurate sample-specific band gap or quasiparticle spectrum.

As to the theory-experiment comparison, strictly speaking the  $\tau_\mathrm{IC}$ value~\cite{Elwell2025CEMS} was measured not in bulk \ce{ThO2}, but in an electrodeposited \ce{^{229}ThO2} film with an estimated thickness of about $10~\mathrm{nm}$. Because the VUV absorption length and the IC-electron escape depth are both of order $10~\mathrm{nm}$, the measurement primarily probes a near-surface region rather than an ideal bulk crystal. In addition, surface/interface relaxation, residual strain, surface chemistry and possible contaminants in these pioneering measurements may further modify the local electronic structure relevant for IC. Furthermore, the present calculation retains only the nuclear \(M1\) contribution to
the IC rate.  For an ideal Th site in cubic ThO$_2$, the
electric-quadrupole (\(E2\)) contribution vanishes by symmetry
\cite{Elwell2025CEMS}.  In an experimental sample, however, defects,
surfaces, strain, or other local distortions may lower the site symmetry
and thereby activate an \(E2\) contribution to the IC decay.  Quantifying
such effects would require calculations for symmetry-broken local
environments and lies beyond the scope of the present bulk-crystal treatment. Such an analysis is especially challenging as it requires experimental characterization of defects and surface contaminants.
A systematic inclusion of these enumerated effects would provide a more qualitatively accurate description of the IC rate, and represents an important direction for future work. 

\begin{table*}[th!]
\caption{
Convergence of the calculated IC rate and lifetime with respect to the
\(k\)-mesh size. Results are shown for the
lowest and highest experimental ThO\(_2\) band-gap values considered.
The corresponding scissor shifts are
\(\Delta_{\rm sc}=0.89~{\rm eV}\) for \(E_\mathrm{gap}=5.20~{\rm eV}\) and
\(\Delta_{\rm sc}=2.69~{\rm eV}\) for \(E_\mathrm{gap}=7.00~{\rm eV}\), using
\(E_\mathrm{gap}^{\rm PBE}=4.31~{\rm eV}\).  Here the smearing width \(\sigma=0.10~{\rm eV}\). 
}
\label{tab:icrate_scissors_bandgaps_sigma010}
\begin{ruledtabular}
\begin{tabular}{lcccc}
\(k\)-mesh &
\multicolumn{2}{c}{\(E_\mathrm{gap}=5.20~{\rm eV}\)} &
\multicolumn{2}{c}{\(E_\mathrm{gap}=7.00~{\rm eV}\)} \\
\cline{2-3}\cline{4-5}
&
\(\Gamma_{\rm IC}\) \(({\rm s}^{-1})\) &
\(\tau_{\rm IC}\) \((\mu{\rm s})\) &
\(\Gamma_{\rm IC}\) \(({\rm s}^{-1})\) &
\(\tau_{\rm IC}\) \((\mu{\rm s})\) \\
\hline
\(10\times10\times10\) & \(9.15\times10^{5}\) & 1.093
& \(5.79\times10^{4}\) & 17.28 \\
\(12\times12\times12\) & \(9.39\times10^{5}\) & 1.065
& \(6.09\times10^{4}\) & 16.43 \\
\(16\times16\times16\) & \(9.07\times10^{5}\) & 1.103
& \(6.06\times10^{4}\) & 16.50 \\
\(32\times32\times32\) & \(8.85\times10^{5}\) & 1.131
& \(6.07\times10^{4}\) & 16.47 \\
\end{tabular}
\end{ruledtabular}
\end{table*}

To summarize the discussion so far, the uncertainty in $\tau_{\rm IC}$ mainly arises from the ThO$_2$ electronic
structure. The dominant
electronic sensitivity comes from the uncertainty in the experimental band gap, which determines the resonant, matrix-element-weighted interband spectrum at
$\hbar\omega_{\rm nuc}$. The tested $k$-mesh and smearing-width sensitivities are comparatively small. 
% Compared to the accuracy of electronic-structure calculations, we find a negligible sensitivity to Bohr-Weisskopf effect in our calculations. 

\subsection{Systematic uncertainties}
Beyond the uncertainties discussed above, the absolute band-to-band IC rate may also be
affected by the uncertainties in the reduced transition probabilities $B(M1)$, the electron-hole correlations beyond the independent-particle
approximation employed in our calculation, quasiparticle corrections beyond the rigid scissor shift, and the Pauli-reduced treatment of the hyperfine operator.

\begin{figure}[h!]
\centering
\includegraphics[
    width=1.0\columnwidth %,
    % height=0.58\textheight,
    % keepaspectratio
]{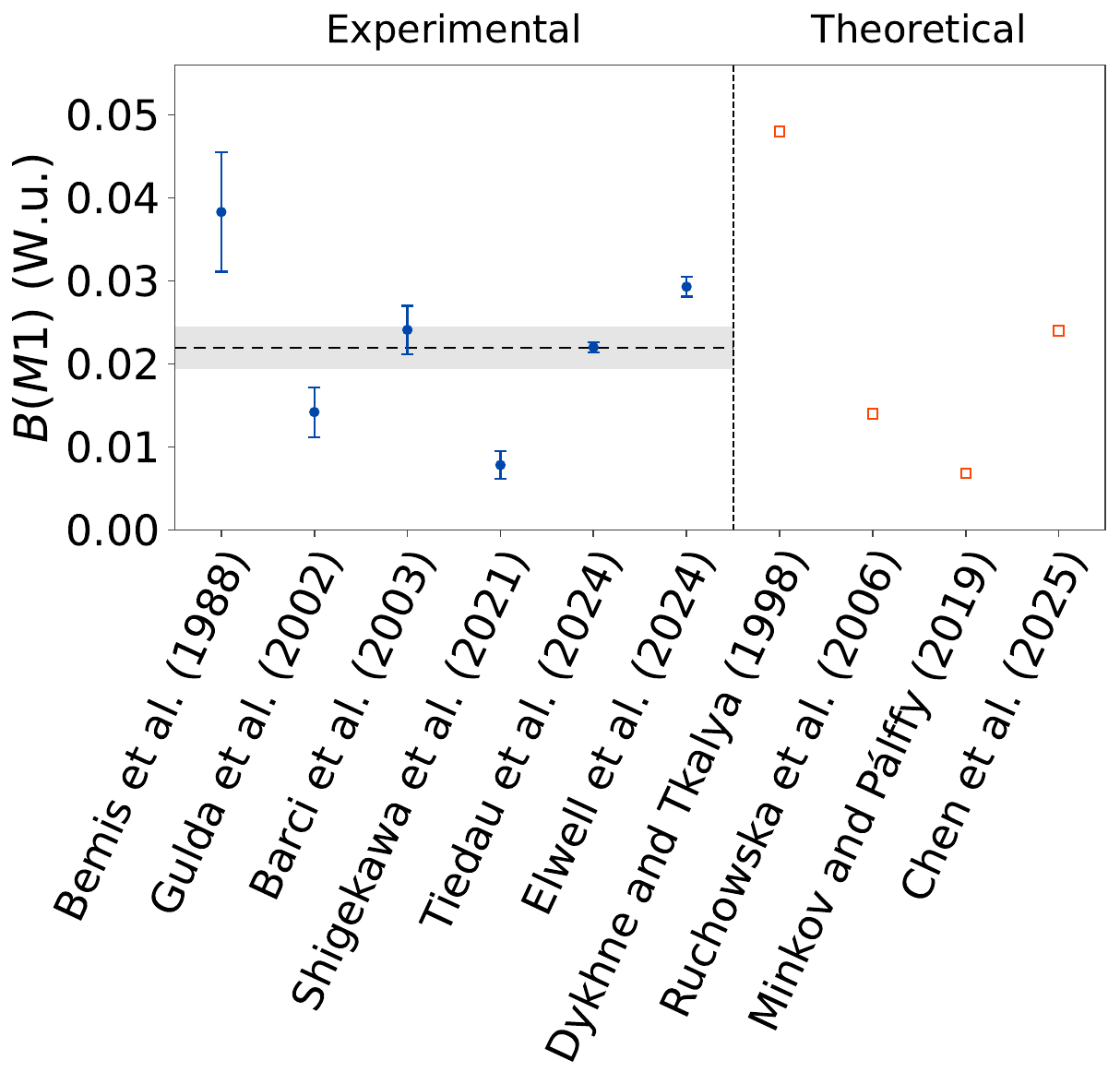}
\caption{
Comparison of selected experimental and theoretical
$B(M1)$ values for the $^{229\mathrm{m}}\mathrm{Th}$
$(3/2^{+}\rightarrow 5/2^{+})$ isomeric transition.
Filled circles denote selected experimental or experimentally inferred values from
Refs.~\cite{Bemis1988Th229,Gulda2002Th229,Barci2003Th229,
Shigekawa2021Th229BM1,Tiedau2024Laser,Elwell2024PRL}.
Open squares denote published theoretical or model-based values from
Refs.~\cite{Dykhne1998Th229Isomer,Ruchowska2006Th229,
Minkov2019Th229MDM,Chen2025Th229PSM}.
The horizontal dashed line indicates the inverse-variance weighted mean of
the selected experimental/inferred values,
$\overline{B(M1)}_{\mathrm{exp,w}}=0.0219(24)~\mathrm{W.u.}$
}
\label{fig:BM1_experimental_theoretical_comparison}
\end{figure}

The IC rate~\eqref{eq:GammaIC_in_terms_of_Sel_BM1} is proportional to the  reduced transition probabilities $B(M1)$ of the nuclear transition and the uncertainties in $B(M1)$ directly translate into those in $\tau_{\rm IC}$.
Literature values of $^{229{\rm m}}$Th $B(M1)$ are
compared in Fig.~\ref{fig:BM1_experimental_theoretical_comparison}.
The experimental and theoretical values span a noticeable range. In
our calculation we used the recent~\cite{Tiedau2024Laser} \(B(M1)=0.0220(6)\,\mathrm{W.u.}\) as the reference value, which
is close to the weighted mean of the experimental
values, $\overline{B(M1)}_{\mathrm{exp,w}}=0.0219(24)$ W.u. This mean value has an 11\% error bar that would imply identical uncertainty in the computed $\tau_{\rm IC}$ values. 

Moreover, the present implementation uses the Pauli-reduced M1 hyperfine operator together with scalar-relativistic plus spin-orbit Bloch spinors. For Th, $Z=90$, so $(\alpha Z)^2 \approx 0.43$ is not a small parameter, and a fully relativistic four-component treatment may be important, especially for the Th-projected $p_{1/2}-p_{1/2}$ channel that is one of the dominant ones in the semi-empirical estimates of Ref.~\cite{Elwell2025CEMS}. Relativistic contraction generally increases the electron density near the nucleus, and therefore can enhance the HFI matrix elements and the corresponding IC rate. Recent LAPW calculations
for \ce{ThO2} by ~\citet{Nikolaev2025} have also emphasized
the importance of relativistic effects for the near-nuclear electronic structure. Fully relativistic implementations for periodic materials remain computationally demanding, however, and are not yet routinely available for this class of calculations.

The independent-particle (IP) treatment employed in our calculations neglects electron-hole correlations in the
final electronic excitation. In an insulating host, however, the promoted
electron and valence hole interact through the screened Coulomb interaction
and may form bound or resonant excitonic states. These correlations can shift
the relevant excitation energies and redistribute spectral weight among the
underlying Bloch transitions. Such effects are routinely described in optical
spectra using the Bethe-Salpeter equation (BSE)
\cite{Albrecht1998,Rohlfing2000,Onida2002}.

Excitonic correlations can affect the IC rate in two ways.  First, the
resonance condition is set by the correlated electronic excitation energy,
rather than by the independent-particle band-to-band energy alone.  Excitonic effects could increase the IC rate by adding additional transitions at \(\hbar\omega_{\rm nuc}\) with significant thorium orbital contributions, or decrease it by redistributing thorium-centered transitions to energies lower than \(\hbar\omega_{\rm nuc}\). Second, because an
exciton is a coherent superposition of electron-hole configurations, the
effective local M1 hyperfine matrix element can differ from the
corresponding independent-particle estimate.  Depending on the phases and
magnitudes of the contributing Bloch-transition matrix elements, this
mixing may either increase or decrease the total IC rate.

We have not included excitonic correlations explicitly in the present
work.  A consistent treatment would require solving the BSE for the
correlated electron-hole eigenstates of the host, or of the relevant
defect-containing crystal, and then contracting the resulting excitonic
amplitudes with the same local Th-centered hyperfine matrix elements
evaluated here.  To our knowledge, such a calculation has not yet been
reported for solid-state \(^{229}\)Th nuclear-clock systems.

As a qualitative estimate of the possible size of electron-hole correlation effects
in the electronic spectrum near the nuclear transition energy, we notice that similar to the IC rate decomposition~\eqref{Eq:gammaIC-JDOS},
the electric-dipole absorption coefficient $\alpha(\omega)$ for interband transitions driven by a laser of frequency $\omega$  is proportional to JDOS \(G_{cv}(\hbar\omega)\)~\cite{callaway1991Book}. Thereby, we 
define the absorption-renormalization factor
\begin{equation}
R_{\alpha}^{\rm p,q}(\omega_{\rm nuc}) =
\frac{\alpha_{\rm p}(\omega_{\rm nuc})}
{\alpha_{\rm q}(\omega_{\rm nuc})} \sim \frac{\Gamma_\mathrm{IC,p}}{\Gamma_\mathrm{IC,q}},
\end{equation}
where \(\alpha_{\rm p}\) and $\Gamma_\mathrm{IC,p}$ are the absorption
coefficient and the IC rate computed with electronic structure method $\rm p$. 
For example, $R_{\alpha}^{\rm BSE-IP,BSE-TD}(\omega_{\rm nuc})$ is the renormalization factor between absorption coefficients from the BSE with and without the independent particle approximation.
In both cases, the one-electron energies are computed with $G_0W_0$ to correct for underestimation of the unoccupied energies by DFT.
``BSE-TD'' denotes that the Tamm-Dancoff approximation was used in the full (\textit{i.e.} excitonic) BSE calculation.

At \(\hbar\omega_{\rm nuc}=8.35~{\rm eV}\), we find $\alpha_{\rm PBE-IP}=1.0666\times10^{6}~{\rm cm}^{-1}$, 
$\alpha_{\rm BSE-IP}=3.0141\times10^{6}~{\rm cm}^{-1}$, and $\alpha_{\rm BSE-TD}=1.1978\times10^{6}~{\rm cm}^{-1}$.
The factor $R_{\alpha}^{\rm BSE-IP,BSE-TD}(8.35~{\rm eV})=2.52$ estimates the influence of excitonic effects on the IC lifetime.
The exclusion of excitons from the BSE calculation changes the optical spectral weight at the nuclear transition energy by a factor of 2-3.
This suggests that excitonic redistribution of the electronic
spectrum near \(\omega_{\rm nuc}\) is significant in ThO$_2$.
The absorption spectra in Fig.~\ref{fig:absorption spectra} support that suggestion and show that the energy window in which excitonic effects are important is  $\sim$6-10 eV, consistent with the band gap of 6.20 eV.
Interestingly, $R_{\alpha}^{\rm PBE-IP,BSE-TD}(8.35~{\rm eV})=0.89$, meaning that TDDFT-IP agrees with BSE-TD much more closely than BSE-IP, probably due to fortuitous error cancellation. This suggests possible error cancellation when using PBE orbitals to calculate the IC rate. 
The PBE-IP absorption coefficient was computed with a scissor shift to match the PBE band gap to the $G_0W_0$ band gap of 6.20 eV.
The spectra are plotted in Appendix \ref{optical spectra appendix}.
Note that these spectra were calculated with VASP; computational details are given in Appendix \ref{optical spectra appendix}.

We emphasize that \(R_{\alpha}\) is only an order-of-magnitude diagnostic and should not be interpreted as a multiplicative correction to \(\Gamma_{\rm IC}\).
Optical absorption probes the long-wavelength electric-dipole response, whereas the IC rate is governed by the local Th-centered magnetic-hyperfine operator.
A quantitative excitonic correction to the IC lifetime would therefore require an explicit BSE-based evaluation of the hyperfine-coupled excitonic matrix elements.
 
\subsection{Qualitative insights}

To facilitate further discussion of our numerical results for IC rates, we follow the common approximation in computing the electric-dipole absorption rates in the interband transitions and separate out the joint density of states (JDOS) from matrix elements~\cite{callaway1991Book}. To this end, for each VB-CB combination we define a surface $S_E$ of constant energy in  $\mathbf{k}$-space through an implicit relation $\varepsilon_{c\mathbf{k}} - \varepsilon_{v\mathbf{k}} =  \hbar \omega_\text{nuc}$.  If the HFI matrix element remains reasonably constant on $S_E$, the rate simplifies to~\cite{Elwell2025CEMS}
\begin{align}
\Gamma_\text{IC} =
  \frac{2 \pi}{\hbar }  \,  
 \left( \frac{1}{[I_e]} \sum_{M_gM_e} 
\overline{|{W^{ge}_{c v}(\mathbf{k})}  |^2} \right)
G_{cv}(\hbar\omega_\text{nuc}) \,,
\label{Eq:gammaIC-JDOS}
 \end{align}
where overline denotes averaging over the surfaces of constant energy $S_E$ and JDOS is defined as
\begin{align}\label{Eq:JDOS}  
 G_{cv}(E) &=   
\frac{\Omega}{(2 \pi)^3} \sum_{cv} \int_\text{BZ} d^3 k \, \delta(\varepsilon_{c\mathbf{k}} - \varepsilon_{v\mathbf{k}}- E)  \nonumber\\
& \approx  \sum_{cv} \sum_k w_k \delta(\varepsilon_{c\mathbf{k}} - \varepsilon_{v\mathbf{k}}- E)\,.
\end{align}

For a transition between two parabolic bands with extrema at $\mathbf{k}=0$, separated by a gap $E_\mathrm{gap}$ at $\mathbf{k}=0$, cf. Fig.~\ref{fig:band_to_band_IC_schematic}, JDOS scales as~\cite{callaway1991Book}
$
   G_{cv}(\hbar \omega_\mathrm{nuc}) \propto \left(\hbar \omega_\mathrm{nuc}-E_\mathrm{gap}\right)^{1/2}
$. This implies that $\tau_\mathrm{IC} \propto \left(\hbar \omega_\mathrm{nuc}-E_\mathrm{gap}\right)^{-1/2}$. 
%Such dependence is in qualitative agreement with the computed IC lifetime vs $E_\mathrm{gap}$ trend in Fig.~\ref{fig:scissor_exp_bandgaps}.
This parabolic band approximation reveals that $\tau_{IC}$ value becomes increasingly sensitive to the gap value as $E_\mathrm{gap}$ approaches $\omega_\mathrm{nuc}$, supporting the observed trend in Fig.~\ref{fig:scissor_exp_bandgaps}.

\begin{figure}[t]
    \centering
    \includegraphics[width=0.95\columnwidth]{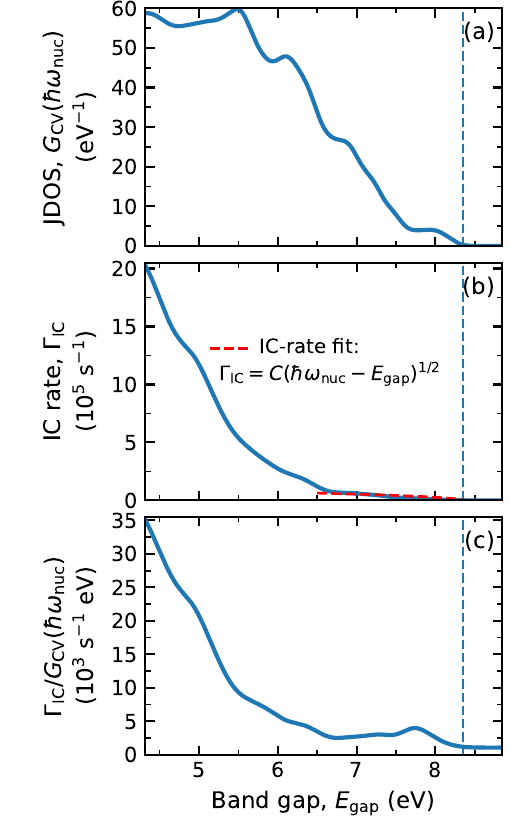}     \caption{
    Dependence of the IC rate on the scissor-shifted band gap \(E_{\mathrm{gap}}\) in \({}^{229}\)ThO\(_2\). Panel (a) shows the joint density of states evaluated at the nuclear transition energy, \(G_{cv}(\hbar\omega_{\mathrm{nuc}})\). Panel (b) shows the calculated IC rate \(\Gamma_{\mathrm{IC}}\), together with a fit to the asymptotic parabolic-band form (dashed red line).
    The coefficient \(C\) is obtained from a least-squares fit to the calculated IC-rate data in the large-gap region \(6.5 \le E_{\mathrm{gap}} < \hbar\omega_{\mathrm{nuc}}\), where the near-threshold square-root scaling is expected to be adequate.
    Panel (c) shows the ratio \(\Gamma_{\mathrm{IC}}/G_{cv}(\hbar\omega_{\mathrm{nuc}})\), which isolates the matrix-element-weighted contribution to the IC rate.
    The vertical dashed line indicates the nuclear transition energy.
    }
    \label{fig:jdos_icratio}
\end{figure}

Fig.~\ref{fig:jdos_icratio} explores the range of validity of the IC rate decomposition~\eqref{Eq:gammaIC-JDOS}.  
The computed PBE JDOS \(G_{cv}(\hbar\omega_{\rm nuc})\) decreases strongly as the scissor-shifted gap increases, see Fig.~\ref{fig:jdos_icratio}(a).
This reduction is expected: when the band gap increases, the number of vertical interband transitions satisfying the resonance condition \(\varepsilon_{c\mathbf{k}}-\varepsilon_{v\mathbf{k}}=\hbar\omega_{\rm nuc}\) is progressively reduced.  
 Eq.~\eqref{Eq:gammaIC-JDOS} shows that \(\Gamma_{\rm IC}\) is proportional to \(G_{cv}(\hbar\omega_{\rm nuc})\) when the matrix element is approximately constant on the constant-energy surface \(S_E\), the IC rate correspondingly decreases with increasing $E_\mathrm{gap}$.
Fig.~\ref{fig:jdos_icratio}(c) shows the ratio \(\Gamma_{\rm IC}/G_{cv}(\hbar\omega_{\rm nuc})\), which removes the leading phase-space contribution of the JDOS. If the decomposition~\eqref{Eq:gammaIC-JDOS} were to remain valid, this ratio would remain flat. It is nearly flat when $E_\mathrm{gap} \gtrsim 7\, \mathrm{eV}$. Here the HFI matrix elements sample states near the valence-band top and conduction-band bottom. For lower band gap values, the matrix elements exhibit much stronger dependence on the band structure. In Fig. \ref{fig:jdos_icratio}(b),  we fit our computed IC rate for $E_\mathrm{gap} > 6.5$ eV to the  expected $\left(\hbar \omega_\mathrm{nuc}-E_\mathrm{gap}\right)^{1/2}$ dependence, showing a reasonable agreement with our expectation, even though the quality of the fit near $E_\mathrm{gap} \approx \omega_\mathrm{nuc}$ may have been affected by the 0.1 eV smearing width. The microscopic origin of the low-gap structure in Fig.~4(c) is
discussed in Appendix~\ref{k_resolved_ic}.

HFI matrix elements depend on the electronic structure in a non-trivial manner. These matrix elements are determined by the overlap of the Th muffin-tin orbitals~\eqref{Eq:Psi-MT-expansion} with the nuclear region. For a given orbital momentum $l$ in Eq.~\eqref{Eq:Psi-MT-expansion}, the square of the HFI matrix element is roughly proportional to the product of Th-projected density of states (PDOS) in the conduction and valence band. Following~\cite{Elwell2025CEMS}, we refer to these contributions as $l-l$ VB-CB channels. PDOS depend on the chemical bond nature of the \ce{ThO2} electronic bands. The electronic structure of fluorite \ce{ThO2} is characteristic of a predominantly ionic oxide with an appreciable degree of Th-O covalency arising from hybridization between O $2p$ and Th $6d$/$5f$ orbitals. The VB is composed primarily of O $2p$ states, while the CB is dominated by unoccupied Th $5f$ states, with Th $6d$ states contributing at somewhat higher energies and hybridizing with the $5f$ manifold~\cite{Prodan2007,Petit2010,Dorado2013,Singh2019}. Consequently, the fundamental band gap of \ce{ThO2} is best viewed as an O $2p \rightarrow$ Th $5f$ charge-transfer gap rather than an intra-atomic \ce{Th} $d$--$d$ or $f$--$f$ excitation. 

\begin{figure}[t]
    \centering
    \includegraphics[width=\linewidth]{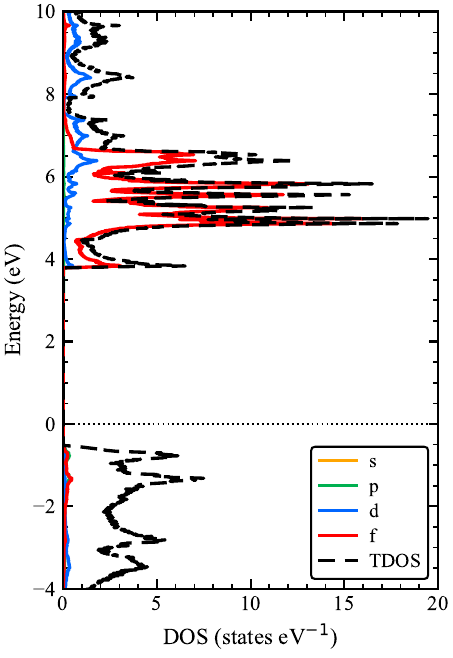}
    \caption{%
    Total density of states (TDOS) and Th-projected density of states
    (PDOS) for \ce{ThO2} computed using the PBE functional. The energy
    is referenced to the Fermi level, indicated by the horizontal dotted
    line at zero energy. Th-PDOS is orbital-resolved, with the $s$, $p$,
    $d$, and $f$ orbital contributions shown in orange, green, blue, and
    red, respectively, while the TDOS is shown by the dashed black curve.
    }
    \label{fig:pdos_elk}
\end{figure}

This point is further illustrated by Fig.~\ref{fig:pdos_elk}, where we plot thorium PDOS in both the valence and conduction bands together with the total density of states (TDOS). While in the VB the TDOS is dominated by oxygen, the finite hybridization between \ce{O} and \ce{Th} orbitals imparts a small but non-negligible Th character to the VB states. This admixture is particularly important for hyperfine-mediated processes, including IC, since the electronic matrix elements depend on the electronic wavefunction amplitude in the vicinity of the Th nucleus, where the dominant contribution originates from the Th component of these otherwise oxygen-dominated valence states. 
As to the conduction band, here the TDOS is dominated by the Th $f$-orbitals and one would have expected that these $f$ states would dominate the IC rate. However,  semi-empirical estimates~\cite{Elwell2025CEMS} show that the $p_{1/2}-p_{1/2}$ CB-VB channel dominates the IC rate due to a larger electronic density of the $p$ states near the nucleus.

Finally, we used the all-electron FP-LAPW+lo method in our calculations.
A widely used alternative for solid-state electronic-structure calculations is the projector-augmented-wave (PAW) formalism~\cite{Blochl1994,Kresse1996}, which has been used in related studies of electronic-bridge processes in \ce{^229Th}-doped crystals~\cite{Nickerson2021,Kirschbaum2025}. PAW should not be viewed as a simple norm-conserving pseudopotential approximation: it introduces a linear transformation between smooth pseudo wavefunctions and reconstructed all-electron wavefunctions, with atom-centered partial waves and projector functions used to recover the rapidly varying near-nuclear behavior. In practical implementations such as VASP, the core states are usually treated within a frozen-core approximation fixed by the PAW dataset, while the valence all-electron character is reconstructed inside the augmentation spheres. This reconstruction has been shown to be sufficiently accurate for many hyperfine-sensitive observables. For example, \citet{Petrilli1998PAW_EFG} applied PAW to electric-field-gradient (EFG) calculations and benchmarked the results against experiment and all-electron LAPW calculations. More recent work has further examined the predictive accuracy of VASP PAW datasets for nuclear-quadrupole-resonance and EFG calculations, finding that the accuracy can depend appreciably on the choice of PAW potential, semicore treatment, and basis-set convergence~\cite{Ansari2019,Choudhary2020EFGDatabase}. However, for the present IC computations, a PAW implementation would require the
off-diagonal valence-to-conduction hyperfine matrix elements to be evaluated,
including the appropriate all-electron PAW augmentation contributions in
the near-nuclear region. Moreover, our treatment
includes a finite nuclear magnetization distribution in the M1 operator,
which would have to be implemented explicitly within the PAW framework.
The resulting IC rate would then need to be checked for convergence with
respect to the PAW dataset, semicore treatment, projector basis, and plane-wave cutoff.

For the present IC problem, the relevant matrix elements probe the electronic wavefunctions in the immediate vicinity of the Th nucleus and are therefore sensitive to the same near-nuclear reconstruction issues that arise in hyperfine and EFG calculations. \citet{Nickerson2021} reconstructed the all-electron wavefunction from PAW calculations performed with VASP and estimated that the resulting uncertainty in the electronic-bridge matrix elements did not exceed about $3\%$. This indicates that carefully reconstructed PAW calculations can provide quantitatively useful near-nuclear electronic information. Nevertheless, the FP-LAPW+lo formulation used here offers a complementary route in which the all-electron wavefunctions are obtained directly in a full-potential basis, without relying on PAW dataset transferability or frozen-core reconstruction for the Th site. This is particularly advantageous for evaluating localized hyperfine-mediated IC matrix elements in a host-dependent crystal environment.

\section{Conclusion}
\label{Sec:Conclusion}

We have developed an all-electron FP-LAPW+lo treatment of band-to-band internal conversion in \ce{^{229}ThO2} that evaluates the local Th-centered hyperfine matrix elements directly from crystal Bloch spinors. The calculation shows that the internal-conversion lifetime is controlled primarily by the resonant, matrix-element-weighted interband spectrum at the $8.35~\mathrm{eV}$ nuclear transition energy. As the \ce{ThO2} band gap is shifted across the experimentally reported range, the calculated lifetime changes from the few-microsecond scale to values comparable with the measured $12.3(3)~\mu\mathrm{s}$ conversion-electron M\"ossbauer lifetime~\cite{Elwell2025CEMS}. This strong gap dependence highlights the need for sample-specific electronic-structure information when comparing solid-state internal-conversion theory with experiment. The present bulk-crystal calculation also identifies the main directions for improvement: explicit quasiparticle and excitonic treatments of the resonant electronic spectrum, calculations for symmetry-broken surface or defect environments, improved constraints on $B(M1)$, and fully relativistic evaluation of the local hyperfine operator. Together, these extensions would turn the present framework into a quantitative tool for predicting internal-conversion quenching and readout channels in \ce{^{229}Th} solid-state nuclear-clock materials.

Looking ahead, these results identify a route toward nuclear-clock optimization through engineering of the electronic environment of the embedded thorium nucleus. The present calculations show that the band-to-band internal-conversion lifetime is highly sensitive to both the host band gap and the local Th-projected electronic structure. This sensitivity suggests that modest changes in composition, stoichiometry, strain, or surface chemistry could be used to tune the decay lifetime from the present $\sim 10~\mu$s range toward $\gtrsim 100~\mu$s. Such a material would retain the practical advantages of internal-conversion readout and optical-pumping-based state preparation, while substantially reducing lifetime broadening.

This strategy may also provide access to narrower resonances than those attainable in conventional transparent nuclear-clock crystals. Crystals that remain transparent at the \ce{^229Th} transition energy are likely to require fluorine-containing, wide-band-gap environments, such as \ce{CaF2}, \ce{LiSrAlF6}, or \ce{ThF4}. However, the \ce{^19F} nuclear spins introduce magnetic-dipole broadening that is difficult to eliminate. By contrast, a spinless, lower-band-gap thorium compound with a deliberately suppressed, but still usable, internal-conversion channel could combine fast electronic readout with reduced magnetic broadening, providing a potentially superior solid-state platform for high-performance nuclear clocks. The guidance developed here suggests that materials composed of nuclear-spin-free atoms with band gaps just below the $8.35~\mathrm{eV}$ isomer energy are promising candidate hosts. Examples include \ce{CaSO4}, \ce{SrSO4}, \ce{BaSO4}, and \ce{Th(SO4)2}~\cite{Morgan2025polyatomic}, which will be the subject of future work.

\section*{Acknowledgments}
This work was supported in part by U.S. National Science Foundation Grants 
PHY-2513134 and PHY-2412982,
and Army Research Office award W911NF2510172.
This work used Bridges-2 at Pittsburgh Supercomputing Center through allocation PHY230110 from the ACCESS program, which is supported by NSF grants \#2138259, \#2138286, \#2138307, \#2137603, and \#2138296.

This work was supported by the Los Alamos National Laboratory LDRD program
project 20260021DR. Calculations were performed using resources provided by the
Los Alamos National Laboratory Institutional Computing program, including
through the Center for Integrated Nanotechnologies, a DOE BES user facility. Los Alamos National
Laboratory, an affirmative action/equal opportunity employer, is managed by
Triad National Security, LLC, for the National Nuclear Security Administration
of the U.S. Department of Energy under contract 89233218CNA000001.

\clearpage
\appendix

\section{Muffin-tin matrix elements of the hyperfine interaction}
\label{app:angular_reduction}

In this appendix, we carry out  angular reduction of the local
Th-centered M1 hyperfine matrix elements used in Eq.~\eqref{eq:Sel_def}
of the main text.  Inside the
Th muffin-tin sphere, with the origin placed at the Th nucleus, the
two-component Bloch spinor is written as
\begin{equation}
\psi_{n{\bf k}}({\bf r})
=
\sum_{lm\sigma}
R^{n{\bf k}}_{lm\sigma}(r)
Y_{lm}(\hat{\bf r})
\chi_\sigma ,
\qquad
\sigma=\pm \frac12 .
\label{eq:app_spinor_expansion}
\end{equation}
Here \(R^{n{\bf k}}_{lm\sigma}(r)\) are Th muffin-tin radial expansion
coefficients, \(Y_{lm}\) are spherical harmonics, and \(\chi_\sigma\) are Pauli spinors.

The local electronic HFI integral for a vertical valence-to-conduction
transition at fixed crystal momentum \({\bf k}\) is
\begin{equation}
I_q(c,v,\mathbf k)
=
\int_{\Omega_{\rm MT}^{\rm Th}}
d^3r\,
\psi^\dagger_{c{\bf k}}({\bf r})
T^{(e)}_q({\bf r})
\psi_{v{\bf k}}({\bf r}),
\label{eq:app_Iq_def}
\end{equation}
where $q=-1,0,+1$ labels spherical components.
The electronic tensor is decomposed as
\begin{equation}
T^{(e)}_q({\bf r})
=
T^{(\mathrm{orb})}_q({\bf r})
+
T^{(\mathrm{sd})}_q({\bf r})
+
T^{(\mathrm{FC})}_q({\bf r}) .
\label{eq:app_T_decomposition}
\end{equation}
For a uniformly magnetized spherical nucleus of radius \(R_N\), the
finite-nucleus operators are
\begin{subequations}
\label{eq:app_finite_T_terms}
\begin{align}
T^{(\mathrm{orb})}_q({\bf r})
&=
\alpha\,
\frac{l_q}{\max(r,R_N)^3},
\label{eq:app_Torb}
\\
T^{(\mathrm{sd})}_q({\bf r})
&=
\alpha\,\frac{g_e}{4}\,
\Theta(r-R_N)
\frac{
3\hat r_q
\left({\boldsymbol\sigma}\cdot\hat{\bf r}\right)
-\sigma_q
}{r^3},
\label{eq:app_Tsd_cart}
\\
T^{(\mathrm{FC})}_q({\bf r})
&=
\alpha\,\frac{2\pi g_e}{3}\,
\frac{\Theta(R_N-r)}{V_N}\,
\sigma_q .
\label{eq:app_TFC}
\end{align}
\end{subequations}
Here \(\alpha\) is the fine-structure constant,
\(g_e=2.0023193\) is the electron g-factor, \(V_N=4\pi R_N^3/3\) is the nuclear volume, and
\({\boldsymbol\sigma}=2{\bf s}\).  Spherical components of a vector are defined by
\begin{equation}
A_0=A_z,
\qquad
A_{\pm1}=\mp\frac{A_x\pm iA_y}{\sqrt2}.
\label{eq:app_spherical_components}
\end{equation}
With this convention,
\[
l_{\pm1}=\mp\frac{l_\pm}{\sqrt2},
\qquad
\sigma_{\pm1}=\mp\frac{\sigma_x\pm i\sigma_y}{\sqrt2}.
\]

The finite nuclear magnetization model modifies only the radial kernels.  Therefore
the angular reduction is the same as for the corresponding point nuclear magnetic-dipole
operator.  We define radial integrals
\begin{equation}
\begin{aligned}
{\cal R}^{(X)}_{l'm'\sigma',lm\sigma}
&=
\int_0^{R_{\rm MT}}
dr\,r^2\,
\left[
R^{c{\bf k}}_{l'm'\sigma}(r')
\right]^*
f^{(X)}(r)
R^{v{\bf k}}_{lm\sigma}(r).
\end{aligned}
\label{eq:app_radial_integral}
\end{equation}
Here \(m\) and \(m'\) denote the valence-state and
the conduction-state magnetic quantum numbers, respectively.
The dependence of \({\cal R}^{(X)}\) on \((c,v,{\bf k})\) is implicit.
Here \(X=\mathrm{orb},\mathrm{sd},\mathrm{FC}\), with
\begin{subequations}
\label{eq:app_radial_kernels}
\begin{align}
f^{(\mathrm{orb})}(r)
&=
\frac{1}{\max(r,R_N)^3},
\label{eq:app_forb}
\\
f^{(\mathrm{sd})}(r)
&=
\frac{\Theta(r-R_N)}{r^3},
\label{eq:app_fsd}
\\
f^{(\mathrm{FC})}(r)
&=
\frac{\Theta(R_N-r)}{V_N}.
\label{eq:app_fFC}
\end{align}
\end{subequations}
Here \(\Theta(x)\) is the Heaviside step function, with
\(\Theta(x)=1\) for \(x>0\) and \(\Theta(x)=0\) for \(x<0\).

The total matrix element is written as the coherent sum
\begin{equation}
I_q(c,v,\mathbf k)
=
I^{(\mathrm{orb})}_{q,cv{\bf k}}
+
I^{(\mathrm{sd})}_{q,cv{\bf k}}
+
I^{(\mathrm{FC})}_{q,cv{\bf k}} .
\label{eq:app_Iq_sum}
\end{equation}
The IC
rate contains
$
\left|I_q(c,v,\mathbf k)\right|^2,
$
which upon expansion yields both 
an incoherent sum of the squared orbital, spin-dipole, and
contact amplitudes and the pairwise interference terms between the three amplitudes.

\subsection{Orbital contribution}

The orbital contribution is
\begin{equation}
I^{(\mathrm{orb})}_{q,cv{\bf k}}
=
\alpha
\sum_{lmm'\sigma}
\langle lm'|l_q|lm\rangle\,
{\cal R}^{(\mathrm{orb})}_{lm'\sigma,lm\sigma}(c,v,{\bf k}) ,
\label{eq:app_Iorb_general}
\end{equation}
where we used the fact that $\langle l'm'|l_q|lm\rangle \propto \delta_{ll'}$. In addition, the orbital operator does not depend on electron spin, fixing $\sigma'=\sigma$.

Using the Wigner--Eckart theorem,
\begin{equation}
\begin{aligned}
I^{(\mathrm{orb})}_{q,cv{\bf k}}
&=
\alpha
\sum_{lmm'\sigma}
(-1)^{l-m'}
\sqrt{l(l+1)(2l+1)}
\\
&\quad\times
\begin{pmatrix}
l&1&l\\
-m'&q&m
\end{pmatrix}
{\cal R}^{(\mathrm{orb})}_{lm'\sigma,lm\sigma}(c,v,{\bf k}) .
\end{aligned}
\label{eq:app_Iorb_WE}
\end{equation}
The \(3j\)-symbol requires $-m'+q+m=0$. We also see that 
the \(l=0\) orbital contribution vanishes identically.

\subsection{Spin-dipole contribution}

To reduce the spin-dipole contribution, we introduce the normalized
spherical harmonic,
\begin{equation}
C^{(2)}_\mu(\hat{\bf r})
=
\sqrt{\frac{4\pi}{5}}\,
Y_{2\mu}(\hat{\bf r}) .
\label{eq:app_C2_def}
\end{equation}

The relevant angular factor can be recast into a rank-1 tensor form,
\begin{equation}
3\hat r_q
\left({\boldsymbol\sigma}\cdot\hat{\bf r}\right)
-
\sigma_q
=
-\sqrt{10}
\left[
C^{(2)}(\hat{\bf r})
\otimes
\sigma^{(1)}
\right]^{(1)}_q \,,
\label{eq:app_sd_identity}
\end{equation}
where
the coupled rank-1 tensor is
\begin{equation}
\left[
C^{(2)}
\otimes
\sigma^{(1)}
\right]^{(1)}_q
=
\sum_{\mu\lambda}
\langle 2\mu,1\lambda|1q\rangle
C^{(2)}_\mu
\sigma_\lambda,
\label{eq:app_coupled_tensor}
\end{equation}
with $\langle 2\mu,1\lambda|1q\rangle$ being the conventional Clebsch-Gordan coefficients.

Therefore,
\begin{equation}
T^{(\mathrm{sd})}_q({\bf r})
=
-\alpha\,\frac{g_e\sqrt{10}}{4}\,
\frac{\Theta(r-R_N)}{r^3}
\left[
C^{(2)}(\hat{\bf r})
\otimes
\sigma^{(1)}
\right]^{(1)}_q .
\label{eq:app_Tsd_coupled}
\end{equation}
The spin-dipole matrix element then reads
\begin{equation}
\begin{aligned}
I^{(\mathrm{sd})}_{q,cv{\bf k}}
&=
-\alpha\,\frac{g_e\sqrt{10}}{4}
\sum_{lm\sigma}
\sum_{l'm'\sigma'}
\sum_{\mu\lambda}
\langle 2\mu,1\lambda|1q\rangle
\\
&\quad\times
\langle l'm'|C^{(2)}_\mu|lm\rangle
\langle\chi_{\sigma'}|\sigma_\lambda|\chi_\sigma\rangle
{\cal R}^{(\mathrm{sd})}_{l'm'\sigma',lm\sigma}(c,v,{\bf k}) .
\end{aligned}
\label{eq:app_Isd}
\end{equation}
Using the Wigner-Eckart theorem and reduced matrix elements of $C^{(2)}$,
\begin{equation}
\begin{aligned}
\langle l'm'|C^{(2)}_\mu|lm\rangle
&=
(-1)^{m'}
\sqrt{(2l'+1)(2l+1)}
\\
&\quad\times
\begin{pmatrix}
l'&2&l\\
0&0&0
\end{pmatrix}
\begin{pmatrix}
l'&2&l\\
-m'&\mu&m
\end{pmatrix}.
\end{aligned}
\label{eq:app_C2_matrix}
\end{equation}
The first \(3j\)-symbol implies
\begin{equation}
|l-2|\le l'\le l+2,\qquad
l+l'= \mathrm{even}.
\label{eq:app_sd_l_selection}
\end{equation}
The second \(3j\)-symbol and the Clebsch--Gordan coefficient in Eq.~\eqref{eq:app_Isd} imply that $m'=m+\mu$ and $\mu+\lambda=q$.

The nonzero matrix elements \(\langle\chi_{\sigma'}|\sigma_\lambda|\chi_\sigma\rangle\) of Pauli matrices are
\begin{subequations}
\label{eq:app_sigma_elements}
\begin{align}
\langle\chi_{\sigma}|\sigma_0|\chi_{\sigma}\rangle
&=
2\sigma,
\label{eq:app_sigma_0}
\\
\langle\chi_{1/2}|\sigma_{+1}|\chi_{-1/2}\rangle
&=
-\sqrt2,
\label{eq:app_sigma_p}
\\
\langle\chi_{-1/2}|\sigma_{-1}|\chi_{1/2}\rangle
&=
+\sqrt2 ,
\label{eq:app_sigma_m}
\end{align}
\end{subequations}
where we used the $\sigma'=\sigma+\lambda$  selection rule.

\subsection{Fermi-contact contribution}

The finite-volume Fermi-contact contribution is
\begin{equation}
\begin{aligned}
I^{(\mathrm{FC})}_{q,cv{\bf k}}
&=
\alpha\,\frac{2\pi g_e}{3}
\sum_{lm}
\sum_{\sigma\sigma'}
\langle\chi_{\sigma'}|\sigma_q|\chi_\sigma\rangle
{\cal R}^{(\mathrm{FC})}_{lm\sigma',lm\sigma}(c,v,{\bf k}) .
\end{aligned}
\label{eq:app_IFC}
\end{equation}
Since the contact kernel is spherically symmetric, the orbital angular integral is diagonal in $l$ and $m$, while the spin part gives the selection rule $\sigma'=\sigma+q$. Matrix elements of Pauli matrices are given by Eq.~\eqref{eq:app_sigma_elements}.

Since we use the finite-volume kernel
\(\Theta(R_N-r)/V_N\), the contact term is not mathematically restricted
to \(l=0\).  Higher-\(l\) contributions are, however, strongly suppressed
by the regular near-origin behavior
\[
R^{n{\bf k}}_{lm\sigma}(r)\propto r^l,
\qquad
r\rightarrow 0 .
\]
In the point-magnetic-moment limit,
\begin{equation}
\frac{\Theta(R_N-r)}{V_N}
\longrightarrow
\delta^{(3)}({\bf r}),
\label{eq:app_delta_limit}
\end{equation}
and only the \(l=0\) partial wave contributes.  The corresponding
point-contact expression is
\begin{equation}
\begin{aligned}
I^{(\mathrm{FC,pt})}_{q,cv{\bf k}}
&=
\alpha\,\frac{g_e}{6}
\sum_{\sigma\sigma'}
\left[
R^{c{\bf k}}_{00\sigma'}(0)
\right]^*
R^{v{\bf k}}_{00\sigma}(0)
\\
&\quad\times
\langle\chi_{\sigma'}|\sigma_q|\chi_\sigma\rangle \,,
\end{aligned}
\label{eq:app_IFC_point}
\end{equation}
with matrix elements of Pauli matrices given by Eq.~\eqref{eq:app_sigma_elements}.

\section{Convergence study}
\label{Sec:Numerical}

The electronic structure was computed with a locally modified version of
the Elk all-electron full-potential linearized augmented-plane-wave
(FP-LAPW+lo) code, version \texttt{Elk-10.5.16}~\cite{ElkCode,ElkManual}, except for calculations of optical absorption spectra (details below).
Exchange and correlation were treated within density-functional theory
using the PBE functional, \texttt{xctype=20}.  All calculations were
performed for the primitive three-atom fluorite ThO$_2$ cell.  Spin-orbit
coupling was included explicitly using \texttt{spinorb=.true.}; in Elk this
adds the \(\boldsymbol{\sigma}\cdot\mathbf{L}\) term to the
second-variational Hamiltonian.  The resulting second-variational states
are two-component spinors.  No symmetry-breaking magnetic field or ordered
magnetic moment was imposed.  The all-electron treatment is essential in
the present problem because the IC matrix elements are dominated by the
near-nuclear electronic amplitude at the Th site.

The IC-specific implementation was built on top of converged Elk
ground-state wave functions and evaluates the relevant local matrix
elements from reconstructed Th-centered muffin-tin spinors.  Where
indicated, a rigid scissor correction was applied in the IC post-processing
to states above the Fermi level in order to align the calculated gap with
experimental estimates.  The corresponding unshifted PBE gaps used in the
analysis are reported separately in the convergence table. For the production calculations we used the \texttt{highq} preset of
\texttt{Elk-10.5.16}, while setting the \(k\)-point meshes explicitly for
each convergence run.  In Elk, \texttt{highq} changes default parameters to
improve convergence; these settings can be overridden by subsequent input
blocks.  The basis parameters are \(\texttt{rgkmax}=8.0\),
\(\texttt{gmaxvr}=16.0\), \(\texttt{lmaxapw}=9\),
\(\texttt{lmaxo}=7\), and \(\texttt{lradstp}=1\). We also used
\(\texttt{lorbcnd=.true.}\) to improve the description of unoccupied
states entering the valence-to-conduction IC transition sum.

For the finite-nucleus calculations reported in this work, we used a
finite nuclear charge distribution, \(\texttt{ptnucl=.false.}\), so that Elk treats the nucleus as a uniformly charged sphere.  The finite
hyperfine-contact model was included in the local IC/hyperfine extension used for the matrix-element evaluation.  The local radial reconstruction inside the Th muffin-tin sphere was carried out with
\(\texttt{lradstp=1}\), so that the finest available radial representation was retained in the near-nuclear region.  The input employed
\(\texttt{rmtscf=1}\) together with \(\texttt{mrmtav=1}\); thus no
additional global scaling of the species-file muffin-tin radii was applied, although Elk's standard radius-averaging procedure remained active.  The
resulting muffin-tin radii in the production runs are \(2.5571\) bohr for
Th and \(1.9888\) bohr for O.  These choices were kept throughout because
the IC matrix elements are especially sensitive to the near-origin spinor
amplitude and therefore to the quality of the radial representation in the
Th muffin-tin.

\textit{Muffin-tin-radius sensitivity.---}
To examine the effect of restricting the hyperfine matrix-element
integration to the Th muffin-tin sphere, we repeated the IC calculations
using
$R_{\mathrm{MT}}^{\mathrm{Th}}=2.5600$, $2.5745$, $2.5912$, and
$2.6072$~bohr, while keeping the crystal structure and all other
numerical parameters fixed. Taking the $2.5600$~bohr calculation as the
reference, the maximum absolute changes in $\Gamma_{\mathrm{IC}}$ over
the nominal experimental band-gap interval
$5.2\leq E_g\leq7.0$~eV are $0.24\%$, $0.46\%$, and $0.68\%$ for
$R_{\mathrm{MT}}^{\mathrm{Th}}=2.5745$, $2.5912$, and $2.6072$~bohr,
respectively. Over the complete scissor-shift range considered, the
maximum difference is $0.77\%$. Thus, varying the Th muffin-tin radius
over this interval changes the calculated IC rate by less than $1\%$.
This result demonstrates that the reported rate is numerically
insensitive to the position of the muffin-tin boundary. It does not,
however, constitute a direct calculation of the complete interstitial
matrix element. We qualitatively expect that the contributions from extending the radial integrals to the interstitial region are suppressed owing to the $1/r^3$ dependence of the orbital and spin-dipole operators.
Interstitial region does not contribute to the Fermi contact terms.

{\em Numerical convergence of the IC lifetime -- }
We tested the stability of the result with respect to the \(k\)-point mesh,
the radial resolution inside the Th muffin-tin sphere, the number of empty
states included in the transition sum, the Elk quality settings, and the
broadening used for the delta function.

The ground-state electronic structure is well converged over the mesh
sequence used in the IC calculation.  Table~\ref{tab:bandgap_totalenergy_kmesh}
shows the convergence of the unshifted PBE Kohn--Sham band gap and total
energy of primitive-cell ThO$_2$.  This confirms that the underlying
ground-state band structure is converged before the scissor correction is
applied.

\begin{table}[t]
\caption{Convergence of the unshifted PBE Kohn--Sham band gap $E_\mathrm{gap}$ and total energy 
$E_\mathrm{tot}$ of primitive-cell \ce{ThO2} with respect to the $k$-point mesh. No scissor-shift correction has been applied.}
\label{tab:bandgap_totalenergy_kmesh}
\begin{ruledtabular}
\begin{tabular}{lcc}
\(k\)-mesh & \(E_{\rm gap}\)\(({\rm eV})\) &$E_\mathrm{tot}(\text{Ha})$  \\
\hline
\(10\times10\times10\) & 4.308022 & \(-27021.1631573\) \\
\(12\times12\times12\) & 4.307993 & \(-27021.1631233\) \\
\(16\times16\times16\) & 4.308004 & \(-27021.1631491\) \\
\(32\times32\times32\) & 4.308019 & \(-27021.1631534\) 
\end{tabular}
\end{ruledtabular}
\end{table}

\begin{figure}[h!]
\centering
\includegraphics[width=8.8cm]{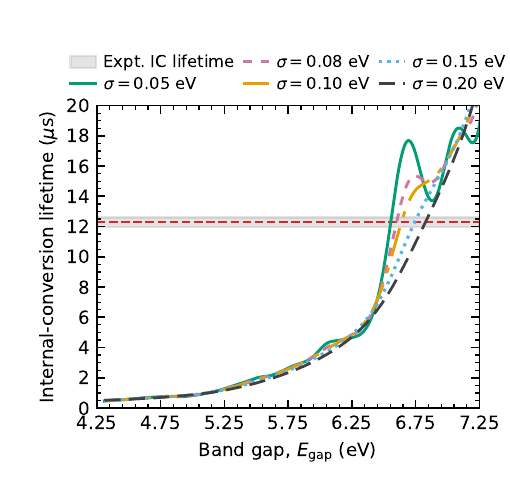}
\caption{
Calculated M1 internal-conversion lifetime in ThO$_2$ as a function of the
scissor-shifted band gap for several values of the smearing width
\(\sigma\). The experimental lifetime is from \citet{Elwell2025CEMS}.
}
\label{fig:sigma_scan_32k}
\end{figure}
For the IC calculation, the energy-conserving delta function was
represented by a finite-width  function.  
In the numerical evaluation of the energy-conserving delta function on a finite
\(k\)-point mesh, we used the Elk Fermi--Dirac derivative broadening,
\begin{equation}
\begin{aligned}
\delta\left(
\varepsilon_{c\mathbf{k}}
-
\varepsilon_{v\mathbf{k}}
-
\hbar\omega_{\mathrm{nuc}}
\right)
&\longrightarrow
\delta_\sigma(x)
\\
&=
\frac{1}{\sigma}
\frac{\exp(-x/\sigma)}
{\left[1+\exp(-x/\sigma)\right]^2},
\end{aligned}
\label{eq:appendix_fd_broadening}
\end{equation}
where
\(x \equiv
\varepsilon_{c\mathbf{k}}
-
\varepsilon_{v\mathbf{k}}
-
\hbar\omega_{\mathrm{nuc}}\).

The smearing width \(\sigma\) is therefore a part of the numerical definition of the
finite-mesh transition sum: it controls how the discrete set of
interband-transition energies samples the continuum limit.  If
\(\sigma\) is too small, the result is sensitive to individual resonant
transitions on the finite \(k\)-mesh; if it is too large, genuine spectral
structure is over-smoothed. The lifetime varies moderately at small
smearing width and becomes comparatively stable for
\(\sigma\simeq 0.08\text{--}0.12~{\rm eV}\), indicating that the chosen
smearing width lies in a numerically stable range. As shown in Fig.~\ref{fig:sigma_scan_32k},  the curves show the 
\(\sigma\) dependence mainly in regions where the resonant transition spectrum varies noticeably. 

The $k$-point convergence pattern of the IC rate is given in
Table~\ref{tab:icrate_scissors_bandgaps_sigma010}. The tested meshes give
stable lifetimes, indicating that the IC rate is well converged with respect
to Brillouin-zone sampling.

\section{Calculated optical absorption spectra}
\label{optical spectra appendix}
Optical absorption spectra were computed with VASP.
The structure of \ce{ThO2} was optimized with VASP 6.4.2 in the conventional cell representation, with the PBE exchange-correlation functional, a 6-6-6 $k$-mesh, and a 500 eV plane-wave cut-off.
This is the same structure that was used in ref. \cite{Elwell2025CEMS}.

\begin{figure}[h!]
    \centering
    \includegraphics[width=0.8\linewidth]{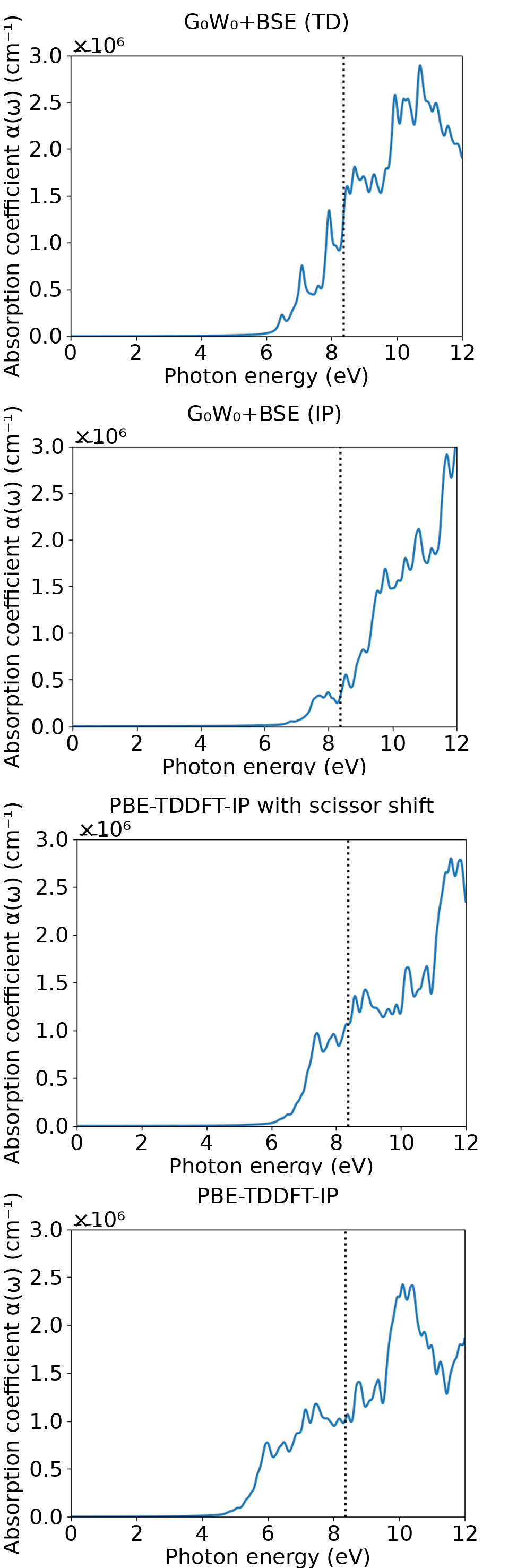}
    \caption{Computed optical absorption spectra of \ce{ThO2} with different methods. The dashed vertical lines mark 8.35 eV, the \ce{^229Th}  transition energy.}
    \label{fig:absorption spectra}
\end{figure}

The spectra were computed with VASP 6.5.1 using two formalisms; time-dependent density functional theory (TDDFT) in the independent particle (IP) approximation, and $G_0W_0$+BSE.
TDDFT calculations were done with and without a scissor correction of 1.4 eV to match the band gap to the $G_0W_0$ value, which is necessary for making a fair comparison between the absorption coefficients. All four spectra are shown in Fig.~\ref{fig:absorption spectra}.
The unshifted TDDFT spectrum is provided for completeness.

The TDDFT calculations used the PBE functional, a 6-6-6 $k$-mesh, and a 500 eV plane wave cutoff.
The $G_0W_0$ calculation used PBE orbitals, an 8-8-8 $k$-mesh, a 500 eV plane-wave cut-off, and 80 frequency grid points (NOMEGA).
All-in-one mode was used such that the number of unoccupied orbitals was set automatically to 1168.
The BSE calculation read 512 bands from the $G_0W_0$ wavefunction and then computed the dielectric function with 8 occupied and 16 unoccupied bands and a maximum pairwise energy difference of 20 eV.
Dielectric and absorption data were processed from the VASP output using VASPKIT.

\section{Microscopic origin of the IC-to-JDOS variation}
\label{k_resolved_ic}

To clarify the variation of $\Gamma_{\mathrm{IC}}/G_{cv}$ in
Fig.~4(c), we analyzed the complete $32\times32\times32$
calculation into contributions from the 897 irreducible
$\mathbf{k}$ points and individual valence-to-conduction
transitions. We used the same nuclear transition energy,
$\hbar\omega_{\mathrm{nuc}}=8.35$~eV, and broadening,
$\sigma=0.10$~eV, as in Fig.~4. The ratio can be written as
\begin{equation}
\frac{\Gamma_{\mathrm{IC}}}{G_{cv}}
\propto
\left\langle
\frac{1}{3}\sum_{q=-1}^{1}
\left|I_q(c,v,\mathbf{k})\right|^2
\right\rangle_{\mathrm{res}},
\label{eq:app_res_average}
\end{equation}
where $\langle\cdots\rangle_{\mathrm{res}}$ denotes an average over
the transitions selected by the broadened resonance condition.
Thus, $\Gamma_{\mathrm{IC}}/G_{cv}$ measures the resonantly averaged
matrix-element strength after the leading transition-phase-space
contribution has been removed.

At each imposed gap, the irreducible $\mathbf{k}$ points were ranked
according to their total contributions to $\Gamma_{\mathrm{IC}}$,
summed over all band pairs. Table~\ref{tab:app_top10_combined}
compares the combined Brillouin-zone weight of the ten leading
points, $\mathcal{K}_{10}(E_{gap})$, with their contributions to
$G_{cv}$ and $\Gamma_{\mathrm{IC}}$.

\begin{table}[h]
\centering
\caption{Combined contributions of the ten irreducible
$\mathbf{k}$ points with the largest contributions to
$\Gamma_{\mathrm{IC}}$ at each imposed band gap. The same
gap-dependent set $\mathcal{K}_{10}(E_{gap})$ is used to evaluate the
$G_{cv}$ and $\Gamma_{\mathrm{IC}}$ percentages.}
\label{tab:app_top10_combined}

\begin{ruledtabular}
\begin{tabular}{cccc}
$E_{gap}$ (eV) &
$\displaystyle\sum_{\mathbf{k}\in\mathcal{K}_{10}} w_{\mathbf{k}}$ &
\multicolumn{2}{c}{Contribution (\%)} \\
& & $G_{cv}$ & $\Gamma_{\mathrm{IC}}$ \\
\hline
4.316 & 0.01172 & 0.919 & 14.616 \\
5.499 & 0.01025 & 1.445 & 27.276 \\
6.771 & 0.01208 & 1.457 & 4.289 \\
7.280 & 0.01465 & 2.235 & 6.526 \\
7.430 & 0.01465 & 2.212 & 6.967 \\
7.744 & 0.01465 & 3.295 & 12.373 \\
\end{tabular}
\end{ruledtabular}
\end{table}

At $E_{gap}=5.499$~eV, the IC rate is distributed nonuniformly over the Brillouin zone. The ten $\mathbf{k}$ points
with the largest IC contributions have a combined Brillouin-zone
integration weight of only $1.025\%$ and account for $1.445\%$ of
$G_{cv}$, but they contribute $27.276\%$ of
$\Gamma_{\mathrm{IC}}$. This disparity shows that their enhanced
contribution originates from unusually large local hyperfine
matrix-element strengths rather than from a larger resonant
transition phase space. Among these ten points, the largest
individual contribution occurs at the fractional reciprocal-lattice
coordinates
$\mathbf{k}=(0.15625,0.09375,0.03125)$ and accounts for
$4.655\%$ of the total IC rate.  Different
$\mathbf{k}$ points and band pairs therefore dominate at different
imposed gaps. This redistribution reduces the resonantly averaged
squared matrix-element strength by approximately a factor of 14
between $E_{gap}=4.316$ and $6.771$~eV.

Moreover, the Th-projected DOS in Fig.~5 provides qualitative
explanation for this behavior. The valence states are predominantly
O $2p$, with a small Th contribution arising from O--Th
hybridization, whereas the conduction manifold contains
energy-dependent Th $5f$ and $6d$ character. Within the
rigid-scissor approximation, the conduction manifold is shifted in
energy without changing the shapes or orbital characters of these
states. Consequently, varying the imposed gap causes the fixed
nuclear transition energy to sample different
valence-to-conduction transitions. Because the hyperfine interaction
probes the electronic wave functions near the Th nucleus, transitions
with similar JDOS contributions can have substantially different IC
matrix elements. Together with the $\mathbf{k}$-resolved analysis,
this shows that the variation of
$\Gamma_{\mathrm{IC}}/G_{cv}$ originates primarily from the changing
resonant sampling of transitions with different Th-local
Fermi-contact matrix-element strengths.

\bibliographystyle{apsrev4-2}
\bibliography{all_cited-ThO2-IC}
\end{document}